\documentclass[journal, 12pt, draftclsnofoot, onecolumn]{IEEEtran}
\IEEEoverridecommandlockouts

\usepackage{amsfonts,amssymb}
\usepackage{ifpdf}
\usepackage{cite}
\usepackage{stfloats}
\usepackage{bm}
\usepackage{color,xcolor}
\usepackage{subfigure}
\ifCLASSINFOpdf
   \usepackage[pdftex]{graphicx}
   \graphicspath{{../pdf/}{../jpeg/}}
   \DeclareGraphicsExtensions{.pdf,.jpeg,.png}
\else
   \usepackage[dvips]{graphicx}

   \graphicspath{{../eps/}}

   \DeclareGraphicsExtensions{.eps}
\fi

\usepackage[cmex10]{amsmath}

\usepackage{algorithm}
\usepackage{algorithmic}
\ifCLASSOPTIONcompsoc
\else
  \usepackage[caption=false,font=footnotesize]{subfig}
\fi

\usepackage{makecell}
\begin{document}
\title{Joint Dynamic Passive Beamforming and Resource Allocation for IRS-Aided Full-Duplex WPCN}
	
\author{Meng~Hua,
	Qingqing~Wu,~\IEEEmembership{Member,~IEEE}
%Luxi~Yang,~\IEEEmembership{Senior Member,~IEEE,}
%Qingqing~Wu,
%Cunhua~Pan,
%Chunguo~Li,~\IEEEmembership{Senior Member,~IEEE,}
%and~A. Lee Swindlehurst,~\IEEEmembership{Fellow,~IEEE}
%%
%%%\thanks{Copyright (c) 2015 IEEE. Personal use of this material is permitted. However, permission to use this material for any other purposes must be obtained from the IEEE by sending a request to pubs-permissions@ieee.org.}
%%\thanks{Manuscript received April   27, 2019; revised July    29, and accepted November  7, 2019. This work was supported by National Natural Science Foundation of China under Grant  61971128,  Grant 61372101, and Grant 61671144, Scientific Research Foundation of Graduate School of Southeast University  under Grand  YBPY1859 and China Scholarship Council (CSC) Scholarship, National High Technology Project of China  under 2015AA01A703,  Cyrus Tang Foundation Endowed Young Scholar Program under SEU-CyrusTang-201801.   The associate editor coordinating the review of this paper and approving it for publication was Kamel Tourki. (\emph{Corresponding author: Luxi Yang}.)}
\thanks{M. Hua and Q. Wu are with the State Key Laboratory of Internet of Things for Smart City, University of Macau, Macao 999078, China (email: menghua@um.edu.mo; qingqingwu@um.edu.mo). }
%\thanks{   L. Yang and C. Li are with the School of Information Science and Engineering, Southeast University, Nanjing 210096, China (e-mail: \{  lxyang, chunguoli\}@seu.edu.cn).}
%\thanks{Q. Wu is with the State Key Laboratory of Internet of Things for Smart City and Department of Electrical and Computer Engineering, University of Macau, Macao 999078 China (email: qingqingwu@um.edu.mo). }
%\thanks{C. Pan is with the School of Electronic Engineering	and Computer Science, Queen Mary University of London, London E1 4NS, U.K. (e-mail: c.pan@qmul.ac.uk).}
%\thanks{A. L. Swindlehurst is with the Center for Pervasive Communications and Computing, University of California at Irvine, Irvine, CA 92697 USA (e-mail: swindle@uci.edu).}
}
\maketitle
\vspace{-1.2cm}
\begin{abstract}
This paper studies intelligent reflecting surface (IRS)-aided full-duplex (FD) wireless-powered communication network (WPCN), where a hybrid access point (HAP) broadcasts energy signals to multiple  devices for their energy harvesting  in the downlink (DL) and meanwhile receives  information  signals  in the uplink (UL) with the help of IRS.   Particularly, we propose three types of IRS beamforming configurations to strike a balance between the system performance and signaling overhead as well as implementation complexity. 
We first propose the \textit{fully dynamic  IRS beamforming}, where the IRS phase-shift vectors vary with each time slot for both DL wireless energy transfer (WET) and UL wireless information transmission (WIT).
 To further reduce  signaling overhead and implementation complexity, we then study two special cases, namely,  \textit{partially dynamic  IRS beamforming} and  \textit{static IRS beamforming}. For the former case,   two different   phase-shift vectors can be exploited for the DL WET and the UL WIT, respectively,  
 whereas for the latter case, the same phase-shift vector needs to be applied for both DL and UL transmissions.
 We aim to maximize the system throughput by jointly optimizing the time allocation, HAP transmit power, and IRS phase shifts  for the above three cases.  Two efficient algorithms based on alternating  optimization  and penalty-based algorithms   are respectively proposed for  both  perfect self-interference cancellation  (SIC) case and imperfect SIC case by applying successive convex approximation  and difference-of-convex  optimization techniques.
 Simulation results  demonstrate the benefits of    IRS  for enhancing the performance of   FD-WPCN, especially with fully dynamic IRS beamforming, and also show that the  IRS-aided FD-WPCN is able to achieve significantly  performance gain compared to its counterpart with half-duplex  when the self-interference (SI) is properly suppressed. 
\end{abstract}

\section{Introduction}
The number of Internet-of-Things (IoT) devices such as temperature sensors, humidity sensors, and  illuminating light sensors,  have  rapidly  skyrocketed recently due to their tremendous  demands required in various application scenarios. Such a massive number of wireless devices thus requires a scalable solution for providing   ubiquitous communication connectivity and  perpetual energy supply in the future. Although the lifetimes of devices can be extended by replacing  and/or recharging the embedded  batteries, it is unsafe and inconvenient especially in a toxic environment and rural areas.  To achieve self-sustainable communication, a variety of wireless technologies has been proposed in the past such as  wireless power transfer  and simultaneous information and power transfer (SWIPT) \cite{lu2015Wireless,wu2017an}. However, the efficiencies of wireless energy  transfer (WET)  and wireless information transmission (WIT)  are severely affected by the distance based path loss and/or the multi-path fading. Although several technologies have been  proposed to overcome this issue, which include  massive multiple-input (MIMO), ultra-dense network (UDN), etc.,
 \cite{zhang2013mimo,wang2018anewlook}, they also face  other challenges in practical implementation  such as high power circuit  consumption and high hardware cost.

To boost the spectral efficiency of wireless systems, full-duplex (FD) transmission  is  a promising technology  to  potentially double the spectral efficiency  if the self-interference (SI) is perfectly cancelled \cite{kim2015Survey}.  Different from the half-duplex (HD) mode that the wireless
node  operates in a time division manner, i.e.,   the transmitter  either   transmits  or  receives signals at one time,    it is able to  transmit signal and receive signal over the same frequency simultaneously for the FD mode. However, due to the simultaneous transmission and reception at the same wireless node, its  receiver antenna will receive the  undesirable signals  transmitted by its nearby transmitter antennas, thus interfering with the desired signal received at the same time.  In fact, the  performance of an FD system may     be even worse than that of an HD system if the SI is not well suppressed.
Fortunately, several SI cancellation (SIC) techniques   have been  proposed in the literature \cite{riihonen2011mitigation,duarte2010full,zhang2015full,zhou2017integrated}, which generally   based on the  analog-domain SIC and the digital-domain SIC.  It was reported in    \cite{zhou2017integrated}   that the art-of-the-state of  SI suppression can be up to $-110~{\rm dB}$   by combing the analog domain (i.e., SIC before  analog-to-digital conversion (ADC) by using analog signal processing techniques) with    digital domain (SIC after ADC by using digital signal processing techniques).  Various works have investigated the  applications of  FD in the different setups, such as  WET and  SWIPT systems \cite{wei2019resource,ju2014optimal,kang2015full,leng2016multi,xu2018Hybrid}. In particular,  the authors in \cite{ju2014optimal}    studied  FD  wireless-powered communication network (WPCN) and aimed at maximizing  the  weighted sum rate over all   users by  jointly  optimizing the   time allocation and  transmit power allocation at a hybrid access point (HAP). The results showed that the FD-WPCN outperforms the HD-WPCN if SI can be suppressed below a certain level. Subsequently,   \cite{kang2015full}   derived closed-form solutions in the FD-WPCN with perfect SIC case.

Recently, intelligent reflecting surfaces (IRSs)  have been proposed as a promising cost-effective solution to improve both spectral- and energy-efficiency of  wireless communication systems\cite{wu2020intelligentarxiv,marco2020smart,WU2020towards,huang2018Achievable,wu2019intelligentxx,hu2021robust}. IRS is  composed of 2-D planar arrays of sub-wavelength metallic or dielectric scatterers, each of which   is able to   independently and smartly induce different reflection amplitudes, phases, and polarization responses on the incident signals and thereby improving   quality of services of users by forming the fine-grained directions of beam towards to the desirable users. In addition,   IRS also has   several other   appealing advantages such as low profile, lightweight, and conformal geometry.  More importantly,   IRS can be composed of large numbers of  dielectric scatterers with a very  limited  size. For example, it was shown in   \cite{tang2021Wireless} that for a large  IRS with $100\times102$ reflecting elements, the electrical sizes are  about $1$ square meter, which is attractive for the practical deployment. As such,  the IRS can be installed on the ceilings to enhance  personal WiFi network  and also can be attached to the facades of buildings to assist the cellular network. Due to the above advantages,   IRS has been exploited  for different applications by  optimizing its reflection coefficients, such as  physical layer security \cite{Intelligent2020guan,zhang2020robust,yu2020robust}, multi-cell cooperation \cite{pan2020multicell,hua2020intelligent,xie2020max},  and  unmanned aerial vehicle communication  \cite{hui2019Reflections,sixian2020robust}. 
While the above works focused on leveraging IRS for enhancing information  transmission,   IRS is also beneficial  for WET and SWIPT \cite{wu2019weighted,pan2019intelligent,li2020joint,wu2020jointActive,wu2021intelligentoverview}. 
By exploiting its large aperture with grained-fine tunable phase shifts,      high passive  beamforming gains can be achieved by the IRS to effectively compensate the end-to-end signal attenuation.  For example,   \cite{wu2020jointActive} 
studied  an IRS-assisted SWIPT system, where a set of IRSs are deployed to assist the downlink (DL) 
WIT and WET from a multi-antenna AP  to multiple  information    and energy users, respectively.  The results showed that the transmission range is significantly enlarged with the help of IRSs.

By far,  there are    only a handful of works paying attention to studying IRS-aided  WPCN \cite{zheng2020Intelligent,zheng2020joint,lyu2021OptimizedEnergy,rezaei2019Secrecy,iqbal2021minimum,wu2021irs}. 
The authors in \cite{zheng2020Intelligent} and \cite{zheng2020joint} studied IRS-aided  WPCN, where an orthogonal protocol for  DL WET and  uplink (UL) WIT was proposed  with the goal of   maximizing the common  and   weighted sum throughput, respectively.
The authors in \cite{lyu2021OptimizedEnergy} studied  an  IRS-empowered WPCN, where the IRS is allowed to  harvest energy from an HAP, and then considered two schemes, namely,  time-switching   and power splitting,    to support   DL WET from the HAP to the distributed users  and  UL WIT from the users to the HAP. In \cite{wu2021irs}, a new optimization framework on dynamic passive beamforming was firstly proposed to compromise the performance and complexity for implementing IRS-aided WPCNs.
 However,  all the above works  focus on the  HD system, which suffers from  a low spectral and energy efficiencies.  

\begin{figure}[!t]
	\centerline{\includegraphics[width=3in]{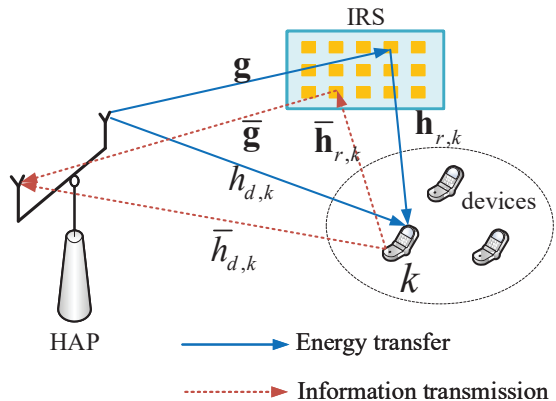}}
	\caption{An IRS-aided FD-WPCN.} \vspace{-0.5cm}\label{fig1}
\end{figure}

In this paper, we study an IRS-aided  FD  WPCN for further improving the system throughput.  As shown in 
Fig.~\ref{fig1}, the HAP  operates in an FD mode with two antennas, which are use for DL WET and UL WIT, respectively. In addition, the IRS is deployed nearby the distributed devices to enhance DL WET from the HAP to the devices and UL WIT from the devices to the HAP simultaneously. 
Note that although there were several works \cite{xu2020resource,shen2020beamforming,cai2020intelligent}
 studied IRS-aided FD systems for some applications,  such as the   cognitive radio system, the  point-to-point system, and  the multi-user multiple-input single-output  system, the above works 
 assumed that  the SI is  either a constant or is  perfectly canceled   by ignoring the practical  quantization
error introduced by the strong SI under the limited dynamic range of the ADC converter. To the best of our knowledge, it is the first work to study  the IRS-aided  FD-WPCN with finite SI. It is worth pointing out that  different from the WPCN without IRS where the channels of all devices remain static within a channel coherence block, we are able to proactively generate optimized artificial time-varying channels by properly designing the IRS reflection coefficients over different time slots within each channel coherence block,  thus  enhancing the multiuser diversity and improving the system throughput. The main contributions  of this paper are summarized as follows.
\begin{itemize}
	\item First, we study  an IRS-aided  FD-WPCN and  propose  three types of IRS beamforming configurations based on how the IRS is allowed to adjust its phase shifts across time.  We first consider the \textit{fully dynamic  IRS beamforming}, where the phase-shift vectors vary with each time slot for DL WET and UL WIT.  To further reduce  signaling overhead and implementation complexity, we then study two special cases, namely,  \textit{partially dynamic  IRS beamforming} and  \textit{static IRS beamforming}. For the former case, there are two different   phase-shift vectors during the whole period with  one  for DL WET and the other  for UL WIT. For the latter case,   DL WET and UL WIT are assumed to adopt the same phase-shift vector. For the above three cases, we
   formulate the corresponding  system throughput maximization problems by jointly optimizing  the time allocation, HAP transmit power, and IRS phase shifts.  It is worth noting that such a thorough study in terms of different dynamic IRS beamforming configurations have not been studied yet in the literature. 
	
	%For comparison, we also consider an optimization problem for HD-WPCN based on  the harvest-then-transmit protocol.
	\item Second, we   consider  fully dynamic  IRS  beamforming optimization with perfect SIC. Since   the formulated problem is non-convex due to the highly coupled optimization variables in the objective function  and non-convex unit-modulus constraints of phase shifts, 	there are  no standard  methods for solving such non-convex optimization problem optimally. To solve this   difficulty,  we propose an efficient  alternating optimization (AO) algorithm by first decomposing the entire variables into two blocks, namely, time allocation and phase shifts, and then optimize each block alternately, until  convergence is achieved. 
	\item Third,  we   consider  fully  dynamic  IRS optimization with imperfect SIC, where the corresponding  problem is more challenging than the former one  due to the additional  HAP transmission power involved in the objective function.  To address this issue, we propose  a novel penalty-based algorithm, which includes a two-layer iteration, i.e.,  an inner layer iteration and an outer  layer iteration. The inner layer solves the penalized optimization problem, while the  outer layer updates the penalty coefficient over iterations to guarantee convergence.  
	
	\item Fourth,  we respectively study partially dynamic IRS beamforming and static IRS beamforming, respectively. Since the formulated problems are    different from  that of  the fully  dynamic  IRS beamforming, we extend  the  AO and penalty-based algorithms to solve them. In  particular, a   difference-of-convex (DC) optimization  method is proposed  to  address  unit-modulus phase-shift constraints, which guarantees to converge to   locally optimal solutions. 

	\item  Finally, simulation results  demonstrate the benefits of the    IRS  used for enhancing the performance of the  FD-WPCN, especially when the  fully dynamic IRS  beamforming is adopted. We  also show that the   IRS-aided FD-WPCN is able to achieve significantly  performance gain compared to the  IRS-aided half-duplex (HD)-WPCN when the SI is well suppressed.
 Furthermore, it is found that  the system with static IRS beamforming    achieves  almost the same   performance as the case with  partially dynamic IRS beamforming   when the HAP has a large transmit power budget.	
\end{itemize}

The rest of this paper is organized as follows. Section II introduces the system model and problem formulations  for FD-WPCN with three types of IRS beamforming configurations. In Section III, we propose an AO based algorithm to solve the problem with  fully dynamic IRS  beamforming  under perfect SIC. In Section IV, we propose a penalty-based  algorithm to solve the problem with  fully dynamic IRS  beamforming  with imperfect SIC. In Section V, we  extend the algorithms to solve the problems with   partially dynamic IRS beamforming and static IRS beamforming, respectively.
 Numerical results are provided in Section VI and  the paper is concluded in Section VII.

\emph{Notations}: Boldface upper-case and lower-case  letter denote matrix and   vector, respectively.  ${\mathbb C}^ {d_1\times d_2}$ stands for the set of  complex $d_1\times d_2$  matrices. For a complex-valued vector $\bf x$, ${\left\| {\bf x} \right\|}$ represents the  Euclidean norm of $\bf x$, ${\rm arg}({\bf x})$ denotes  the phase of   $\bf x$, and ${\rm diag}(\bf x) $ denotes a diagonal matrix whose main diagonal elements are extracted from vector $\bf x$.
For a vector $\bf x$, ${\bf x}^*$ and  ${\bf x}^H$  stand for  its conjugate and  conjugate transpose respectively.   For a square matrix $\bf X$,  ${\rm{Tr}}\left( {\bf{X}} \right)$, $\left\| {\bf{X}} \right\|_2$ and ${\rm{rank}}\left( {\bf{X}} \right)$ respectively  stand for  its trace, Euclidean norm and rank,  while ${\bf{X}} \succeq {\bf{0}}$ indicates that matrix $\bf X$ is positive semi-definite.
A circularly symmetric complex Gaussian random variable $x$ with mean $ \mu$ and variance  $ \sigma^2$ is denoted by ${x} \sim {\cal CN}\left( {{{\mu }},{{\sigma^2 }}} \right)$.  ${\cal O}\left(  \cdot  \right)$ is the big-O computational complexity notation.

\begin{figure}[!t]
	\centerline{\includegraphics[width=6in]{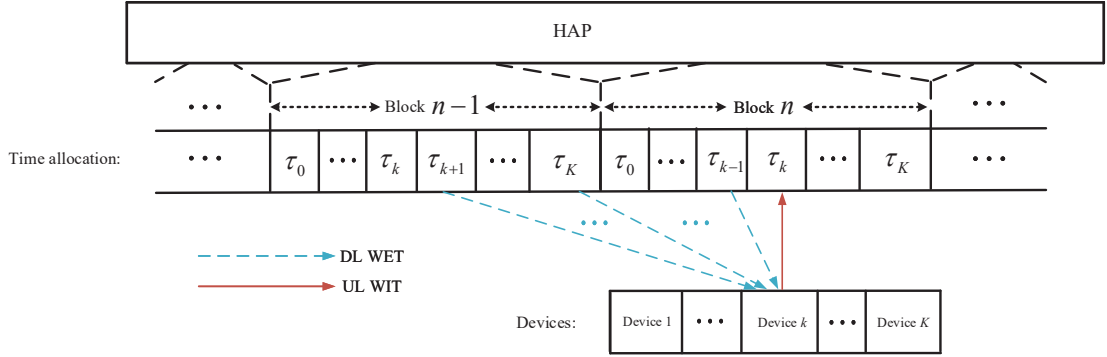}}
	\caption{Transmission protocol  for FD energy harvesting and information transmission.}\vspace{-0.5cm} \label{fig2}
\end{figure}
\section{System Model and Problem Formulation}
\subsection{System Model}
Consider an IRS-aided full-duplex WPCN consisting of an  HAP, $K$ single-antenna devices, and an IRS, as shown in Fig.~\ref{fig1}.  We assume that the HAP operates in the FD mode to enhance the spectral efficiency  and is equipped with two antennas, i.e., a transmitter antenna  and a receiver antenna.   The transmitter antenna   broadcasts energy to the distributed devices in the DL and meanwhile  the receiver antenna   receives the information from  the distributed devices in the UL  simultaneously over the same frequency band. In addition, the distributed devices are assumed to   operate in a time-division HD mode due to their limited signal processing capability, where the devices first  harvest energy in the DL and then transmit information in the UL. 

We consider a quasi-static flat-fading channel  in which the channel state information  remains constant in a channel coherence  frame, but may change in the subsequent frames.
 As shown in Fig.~\ref{fig2},  each channel coherence frame consists of multiple blocks and  each transmission period of one block of interest  denoted by $T$ is divided into $K+1$ time  slots each with duration of $\tau_i,i\in{\cal K}_1$, where ${\cal K}_1=\{0,\dots,K\}$. The $0$th slot is a dedicated time slot used to broadcast  energy to all distributed devices in the DL, while the  $i$th ($i \ne 0$) time slot is used for both DL WET and UL WIT. Since there are two antennas at the HAP, the DL   and UL channels are   different in general. In the DL transmission, denote by  ${\bf{g}} \in {{\mathbb C}^{M \times 1}}$, ${\bf{h}}_{r,k}^H \in {{\mathbb C}^{1 \times M}}$, and ${h_{d,k}} \in {{\mathbb C}^{1 \times 1}}$ the complex equivalent baseband   channel between the HAP  and  the  IRS, between  the IRS  and the $k$th device, and  between  the  HAP and the $k$th device,  $k\in{\cal K}_2$, where ${\cal K}_2=\{1,\dots,K\}$, respectively.  In the UL transmission, denote by  ${\bf{\bar g}}^H \in {{\mathbb C}^{1 \times M}}$, ${\bf{\bar h}}_{r,k} \in {{\mathbb C}^{M \times 1}}$, and ${{\bar h}_{d,k}} \in {{\mathbb C}^{1 \times 1}}$ the complex equivalent baseband   channel between the HAP  and  the IRS, between the IRS  and the $k$th device, and  between  the  HAP and the $k$th device,  $k\in{\cal K}_2$, respectively. In addition, the channel reciprocity is assumed and thus we have   ${{{\bf{\bar h}}}_{r,k}} = {\bf{h}}_{r,k}^*$ for IRS-device links.
 
In this paper, we consider three types of IRS beamforming configurations, namely, fully dynamic IRS, partially dynamic IRS, and static IRS. In the following, we first provide the details for  modeling the  fully dynamic IRS case, and the other two cases are discussed  in Section II-B.
%\footnote{To characterize the fundamental performance limits of IRS-aided FD-WPCN, we first consider dynamic IRS where   the phase shifts  change over each time slot. However,  it  will incur a higher feedback  signaling  overhead between IRS and HAP and large numbers of variables for optimization.
%	To tradeoff between the system performance and the   feedback signaling  overhead, we study  two other cases, namely, static IRS and partially dynamic  IRS, which are discussed in the later sections.} 
\subsubsection{DL WET}
During DL WET, the received signal  by device $k$ in time slot $i$ is given by
\begin{align}
y_{k,i}^d = \sqrt {{P_i}} \left( {{h_{d,k}} + {\bf{h}}_{r,k}^H{{\bf \Theta} _i}{{\bf{g}}}} \right){s_i}+n_{k,i}^d,
\end{align}
where $s_i$  is a pseudo-random signal  which is a prior  known at the HAP satisfying ${\mathbb E}\left\{ {{{\left| {{s_i}} \right|}^2}} \right\} = 1$,  $P_i$ represents the HAP's transmit power at time slot $i$, ${\bf{\Theta }}_i = {\rm{diag}}\left( {{e^{j{ \theta _{i,1}}}}, \cdots ,{e^{j{\theta _{i,M}}}}} \right)$  represents  a diagonal reflection coefficient matrix in  the $i$th time slot, where the reflection amplitude is fixed as $1$ and $\theta _{i,m}$ denotes
 the $m$th reflecting element of the IRS phase shifts  in   the  $i$th time slot, and $n_{k,i}^d \sim {\cal CN} \left( {0,\sigma ^2} \right)$ stands for the additive white Gaussian noise at device $k$ in the $i$th time slot. For ease of exposition, we assume that the $k$th time slot is occupied by device $k$ for UL WIT, $k\in{\cal K}_2$.\footnote{Note that in practice, the device scheduling order for the UL WIT may be affected by many other factors. Thus, in Section VI, we study two scheduling strategies based on the power gain of direct  links, and evaluate their impacts on the system performance. }
 It is worth noting that as in \cite{ju2014optimal}, since we consider a {\it periodic} transmission protocol shown in Fig.~\ref{fig2}, each device $k$ can harvest energy during all time slots except its own transmission time slot $k$.  Particularly, the energy harvested by device $k$ after its own WIT in the previous   transmission block (e.g., $n-1$th) will be used for its WIT in the next transmission block (e.g.,  $n$th). 
Under this protocol, the amount of  harvested energy   by device $k$ during $T$ can be equivalently expressed as\footnote{ 
		The energy harvested from the noise and the received UL WIT signals from other
	devices are assumed to be negligible, since both the noise power and device transmit power are
	much smaller than the transmit power of the HAP in practice \cite{Ju2014Throughput},\cite{wqu2018Spectral}.} 
\begin{align}
{E_k} = {\eta _k}\sum\limits_{i = 0,i \ne k}^K {{P_i}} {\tau _i}{\left| {{h_{d,k}} + {\bf{h}}_{r,k}^H{{\bf{\Theta }}_i}{\bf{g}}} \right|^2},
\end{align}
where  $0\le{\eta _k}\le1$ denotes the  energy conversion coefficient at device $k$. 

On the other hand, when   device  $k$ is scheduled  to transmit in the $k$th time slot, device  $k$ will exhaust  the total harvested energy for  UL WIT. The average transmit power of  device $k$ for  UL WIT  is thus given by 
\begin{align}
{p_k} = {\eta _k}\sum\limits_{i = 0,i \ne k}^K {{P_i}} {\tau _i}{\left| {{h_{d,k}} + {\bf{h}}_{r,k}^H{{\bf{\Theta }}_i}{\bf{g}}} \right|^2}/{\tau _k}, k\in{\cal K}_2.
\end{align}

\subsubsection{UL WIT}
During UL WIT, device  $k$ transmits its own data to the HAP, the  received signal  by the HAP in time slot $k$ is given by  
\begin{align}
y_k^r = \underbrace {\sqrt {{p_k}} \left( {{{\bar h}_{d,k}} + {{{\bf{\bar g}}}^H}{{\bf{\Theta }}_k}{{{\bf{\bar h}}}_{r,k}}} \right){x_k}}_{{\rm{desired}}\;{\rm{signal}}} + \underbrace {\sqrt {{P_k}} \left( {f + {{{\bf{\tilde g}}}^H}{{\bf{\Theta }}_k}{\bf{g}}} \right){s_k}}_{{\mathop{\rm int}} {\rm{erference}}} + n_k^r, \label{up_receiver}
\end{align}
where  ${x_k}$ denotes device  $k$'s  transmit signal, which is assumed to be of zero mean and unit power, i.e., ${\mathbb E}\left\{ {{x_k}} \right\} = 0$ and ${\mathbb E}\left\{ {{{\left| {{x_k}} \right|}^2}} \right\} = 1$,  $f$ represents the effective loopback channel at the  HAP that  satisfies ${\mathbb E}\left\{ {{{\left\| f \right\|}^2}} \right\} = \gamma$, and  $n_{k}^r \sim {\cal CN} \left( {0,\sigma ^2} \right)$ denotes the received additive white Gaussian noise. 
%The desired signal in   \eqref{up_receiver} includes two parts. The first part $\sqrt {{p_k}} {{\bar h}_{d,k}}{x_k}$ denotes the signal   transmitted directly from   the $k$th device  to HAP, the other part $\sqrt {{p_k}} {{{\bf{\bar g}}}^H}{{\bf \Theta} _k}{{{\bf{\bar h}}}_{r,k}}{x_k}$ denotes  the  signal reflected by IRS via  the cascaded  HAP-IRS-device  link. 
It can be observed that the interference term in \eqref{up_receiver}  consists of two parts. The first   $\sqrt {{P_k}} f{s_k}$ denotes the SI resulting from the DL transmission by the HAP  and the second  part $\sqrt {{P_k}} {{{\bf{\tilde g}}}^H}{{\bf \Theta} _k}{{\bf{g}}}{s_k}$ denotes the interference  introduced by the reflection of the DL transmit signal by the IRS. We assume that the IRS is deployed close to the devices and is far from the HAP by considering the  following two  facts. First, when the  IRS is deployed close to the devices,  the double propagation loss of  the  cascaded  HAP-IRS-device  link will be substantially reduced, which is beneficial for improving the signal-to-noise ratio (SNR). Second, when the IRS is deployed far from the HAP, the power of interference  introduced by the reflection  link, i.e., HAP-IRS-HAP link,  will be significantly small compared to that of SI  resulting from the DL transmission. As such, the second part of interference is   neglected in the sequel of this paper.  
\subsubsection{SIC}
In practice, the FD operation is often precluded by the significant quantization error introduced by the strong SI  under the limited
dynamic range of ADC at the receiver. Following \cite{ju2014optimal}, \cite{Day2012Full},   this quantization error after ADC can be modeled as an independent white  Gaussian noise denoted by ${n_{error,k}} \sim CN\left( {0,\beta \sigma _{error}^2} \right),k\in {\cal K}_2$, where    $\beta  \ll 1$ and ${\sigma _{error}^2}$  are given by 
\begin{align}
\sigma _{error,k}^2& = {p_k}{\left| {{{\bar h}_{d,k}} + {{{\bf{\bar g}}}^H}{{\bf \Theta} _k}{{{\bf{\bar h}}}_{r,k}}} \right|^2} + \gamma{P_k} + {\sigma ^2}\notag\\
&\overset{(a)}{\approx} \gamma{P_k},\label{quantization_error}
\end{align}
where $(a)$ holds due to the fact that the power of SI is generally much larger than that of the received  signal from UL devices and the received  noise. Expression \eqref{quantization_error} shows that the power of quantization error is proportional to that of transmit power at the HAP.
In addition, we assume that the perfect loopback channel $f$ is available at the receiver of the HAP via using pilot-based channel estimation methods \cite{Day2012Full}. Therefore,    subtracting SI from \eqref{up_receiver}, the received signal at the HAP can be recast as 
\begin{align}
\tilde y_k^r = \sqrt {{p_k}} \left( {{{\bar h}_{d,k}} + {{{\bf{\bar g}}}^H}{{\bf{\Theta }}_k}{{{\bf{\bar h}}}_{r,k}}} \right){x_k} + {n_{error,k}} + n_k^r.
\end{align}
As a result, the achievable throughput of device $k$ in bits/Hz  in time slot $k$ can be  expressed as
\begin{align}
{R_k} = {\tau _k}{\log _2}\left( {1 + \frac{{{p_k}{{\left| {{{\bar h}_{d,k}} + {{{\bf{\bar g}}}^H}{{\bf \Theta} _k}{{{\bf{\bar h}}}_{r,k}}} \right|}^2}}}{{\Gamma \left( {\beta \gamma {P_k} + {\sigma ^2}} \right)}}} \right),
\end{align}
where $\Gamma $ represents the gap from channel capacity owing to the  practical modulation and coding scheme.

%\subsubsection{Energy causality constraint}
%%When the device  is active to transmit,  we model the static circuit power consumption as a constant,  denoted by $p_{c,k}$, to account for its consumption caused by filters, mixers, and synthesizers.  
% We assume that the device only can use the harvested energy before transmission, as such, we have the following  energy causality constraint given by 
%\begin{align}
% {p_k}{\tau _k} \le {\eta _k}\sum\limits_{i = 0}^{k - 1} {{P_i}} {\tau _i}{\left| {{h_{d,k}} + {\bf{h}}_{r,k}^H{{\bf \Theta} _i}{{\bf{g}}}} \right|^2}.
%\end{align}
%Note that at the  remaining time slots, i.e, from time slots $k+1$ to $K$, user $k$  still  harvest energy from HAP and store it in the battery, but will not  transmit at these time within this block time.
\subsection{Problem Formulation}
 Denote by sets ${\cal M} = \left\{ {1, \ldots ,M} \right\}$, $\tau {\rm{ = }}\left\{ {{\tau _k},k \in {{\cal K}_1}} \right\}$, $P = \left\{ {{P_k},k \in {{\cal K}_1}} \right\}$,  and ${\bf{v}} = \left\{ {{{\bf{v}}_i},i \in {{\cal K}_1}} \right\}$, where  ${\bf{v}}_i^H = {\left[ {{v_{i,1}}, \ldots ,{v_{i,M}}} \right]}$ and 
 ${v_{i,m}} = {e^{j{\theta _{i,m}}}},i \in {{\cal K}_1}, m \in {\cal M}$.
  Let ${{\bf{q}}_k} = {\rm{diag}}\left( {{\bf{h}}_{r,k}^H} \right){\bf{g}}$ and ${{\bf{\bar q}}_k} = {\rm{diag}}\left( {{{{\bf{\bar g}}}^H}} \right){{\bf{\bar h}}_{r,k}}$. 
We thus have 
 ${\left| {{h_{d,k}} + {\bf{h}}_{r,k}^H{{\bf{\Theta }}_i}{\bf{g}}} \right|^2} = {\left| {{h_{d,k}} + {\bf{v}}_i^H{{\bf{q}}_k}} \right|^2}$ and ${\left| {{{\bar h}_{d,k}} + {{{\bf{\bar g}}}^H}{{\bf{\Theta }}_i}{{{\bf{\bar h}}}_{r,k}}} \right|^2} = {\left| {{{\bar h}_{d,k}} + {\bf{v}}_i^H{{{\bf{\bar q}}}_k}} \right|^2}$.
 The  objective of this paper is to   maximize the sum  throughput of the IRS-aided FD-WPCN by jointly optimizing the  time allocation, HAP transmit power, and IRS phase shifts.\footnote{Note that the considered problem formulation  can be readily extended to take into account the fairness among devices  by adding the different weighting factors on each device in the objective function, which does not   affect our proposed algorithms.  As such,  we focus on the  achievable sum throughput maximization problem  instead.} We consider three types of IRS beamforming  configurations,  which are specified as follows. 
\subsubsection{Fully Dynamic  IRS Beamforming}
In this case, the phase-shift vectors change over each time slot during period of $T$.    
Mathematically, the problem is formulated as follows
\begin{subequations} \label{p1}
	\begin{align}
	&\mathop {\max }\limits_{\tau ,P,{\bf{v}}} \sum\limits_{k = 1}^K {{\tau _k}} {\log _2}\left( {1 + \frac{{{\eta _k}\sum\nolimits_{i = 0,i \ne k}^K {{P_i}} {\tau _i}{{\left| {{h_{d,k}} + {\bf{v}}_i^H{{\bf{q}}_k}} \right|}^2}{{\left| {{{\bar h}_{d,k}} + {\bf{v}}_k^H{{{\bf{\bar q}}}_k}} \right|}^2}}}{{\Gamma \left( {\beta \gamma {P_k} + {\sigma ^2}} \right){\tau _k}}}} \right)\label{p1_obj}\\
	&{\rm s.t.}~\sum\limits_{k = 0}^K {{\tau _k}}  \le T,\label{Perfect_problem_const3}\\
	&\qquad{\tau _k} \ge 0,{k \in {{\cal K}_1}},\label{Perfect_problem_const4}\\
	&\qquad 0 \le {P_k} \le {P_{\max }},{k \in {{\cal K}_1}},\label{Perfect_problem_const5}\\
	&\qquad \left\| {{v_{k,m}}} \right\| = 1, m \in {\cal M},{k \in {{\cal K}_1}},\label{Perfect_problem_const2}
	\end{align}
\end{subequations}
where $P_{\max}$ in \eqref{Perfect_problem_const5}  denotes the HAP transmit power budget. 

\subsubsection{Partially  Dynamic  IRS Beamforming}
In this case,   two different  phase-shift vectors can be  allowed to be used for DL WET and UL WIT, which are denoted by  ${\bf v}_d$ and ${\bf v}_u$, respectively. We thus have ${\bf v}_d={\bf v}_0, {\bf v}_u={\bf v}_k, k\in {\cal K}_2$. Define ${{\bf{v}}_t} = \left\{ {{{\bf{v}}_d},{{\bf{v}}_u}} \right\}$. Accordingly, the problem can be formulated as follows 
\begin{subequations} \label{Partially_p}
	\begin{align}
	&\mathop {\max }\limits_{\tau ,P,{{\bf{v}}_t}} \sum\limits_{k = 1}^K {{\tau _k}} {\log _2}\left( {1 + \frac{{{\eta _k}\left( {{P_0}{\tau _0}{{\left| {{h_{d,k}} + {\bf{v}}_d^H{{\bf{q}}_k}} \right|}^2} + \sum\nolimits_{i = 1,i \ne k}^K {{P_i}} {\tau _i}{{\left| {{h_{d,k}} + {\bf{v}}_u^H{{\bf{q}}_k}} \right|}^2}} \right){{\left| {{{\bar h}_{d,k}} + {\bf{v}}_u^H{{{\bf{\bar q}}}_k}} \right|}^2}}}{{\Gamma \left( {\beta \gamma {P_k} + {\sigma ^2}} \right){\tau _k}}}} \right)\\
	&{\rm s.t.}~\left\| {{{\bf v}_{t,m}}} \right\| = 1,m \in {\cal M},t \in \left\{ {u,d} \right\},\\
	&\qquad \eqref{Perfect_problem_const3},\eqref{Perfect_problem_const4},\eqref{Perfect_problem_const5},
	\end{align}
\end{subequations}
where ${{{\bf v}_{t,m}}}$ denotes the $m$th entry of ${{\bf{v}}_t}$. 
\subsubsection{Static IRS Beamforming}
In this case,  we adopt the same   phase-shift vector  for both UL and DL transmission, i.e., $ {\bf v}_0={\bf v}_1=,\dots,={\bf v}_K$. Denote by ${{\bf{v}}_s}$  the phase-shift vector for static IRS.   Accordingly, the problem can be formulated as follows 
\begin{subequations} \label{pStatic}
	\begin{align}
	&\mathop {\max }\limits_{\tau ,P,{{\bf v}_s}} \sum\limits_{k = 1}^K {{\tau _k}} {\log _2}\left( {1 + \frac{{{\eta _k}\left( {\sum\nolimits_{i = 0,i \ne k}^K {{P_i}} {\tau _i}} \right){{\left| {{h_{d,k}} + {\bf{v}}_s^H{{\bf{q}}_k}} \right|}^2}{{\left| {{{\bar h}_{d,k}} + {\bf{v}}_s^H{{{\bf{\bar q}}}_k}} \right|}^2}}}{{\Gamma \left( {\beta \gamma {P_k} + {\sigma ^2}} \right){\tau _k}}}} \right)\label{pStatic_obj}\\
	&{\rm s.t.}~\left\| {{{\bf v}_{s,m}}} \right\| = 1,m \in {\cal M},\label{pstatic_const_1}\\
	&\qquad \eqref{Perfect_problem_const3},\eqref{Perfect_problem_const4},\eqref{Perfect_problem_const5}, 
	\end{align}
\end{subequations}
where ${{{\bf v}_{s,m}}}$ denotes the $m$th entry of ${{\bf{v}}_s}$. 

The above three problems \eqref{p1}, \eqref{Partially_p} and \eqref{pStatic} are all non-convex since the optimization variables are highly coupled in the objective functions, there are no standard methods for solving such non-convex optimization problems optimally in general.  Although it looks like  \eqref{p1} and \eqref{pStatic} have the similar objective function, the hidden structures are   fundamentally different. Specifically,  in \eqref{p1_obj},     ${\sum\nolimits_{i = 0,i \ne k}^K {{P_i}} {\tau _i}{{\left| {{h_{d,k}} + {\bf{v}}_i^H{{\bf{q}}_k}} \right|}^2}}$ and  ${{{\left| {{{\bar h}_{d,k}} + {\bf{v}}_k^H{{{\bf{\bar q}}}_k}} \right|}^2}}$ are not coupled with respect to (w.r.t.) ${{{\bf{v }}_k}}$. While ${{{\left| {{h_{d,k}} + {\bf{v}}_s^H{{\bf{q}}_k}} \right|}^2}}$ and ${{{\left| {{{\bar h}_{d,k}} + {\bf{v}}_s^H{{{\bf{\bar q}}}_k}} \right|}^2}}$  have the same phase-shift vector ${\bf{v }_s}$ in \eqref{pStatic_obj}. Therefore, different algorithms are required.   In the following two sections, we propose to two framework algorithms, namely AO and penalty-based algorithms, based on the successive convex approximation (SCA) and DC optimization techniques to solve problem \eqref{p1}. The extension to solve problems \eqref{Partially_p} and \eqref{pStatic} will be studied in Section V. 
\section{Proposed Solution  For Fully Dynamic IRS Beamforming with Perfect SIC}
In this section, we study the fully  dynamic IRS beamforming with perfect SIC, i.e., ${\gamma= 0}$, which also provides the performance upper bound for the case with imperfect SIC. Problem \eqref{p1} can be simplified to  
\begin{subequations} \label{Perfect_problem}
\begin{align}
	&\mathop {\max }\limits_{P,\tau ,{\bf{v}}} \sum\limits_{k = 1}^K {{\tau _k}} {\log _2}\left( {1 + \frac{{{\eta _k}\sum\nolimits_{i = 0,i \ne k}^K {{P_i}} {\tau _i}{{\left| {{h_{d,k}} + {\bf{v}}_i^H{{\bf{q}}_k}} \right|}^2}{{\left| {{{\bar h}_{d,k}} + {\bf{v}}_k^H{{{\bf{\bar q}}}_k}} \right|}^2}}}{{\Gamma {\sigma ^2}{\tau _k}}}} \right)\label{Perfect_problem_obj}\\
	&{\rm s.t.}~\eqref{Perfect_problem_const3},\eqref{Perfect_problem_const4},\eqref{Perfect_problem_const5}, \eqref{Perfect_problem_const2}.
\end{align}
\end{subequations}
Problem \eqref{Perfect_problem} is   a  non-convex optimization problem  since   the optimization  variables are coupled in \eqref{Perfect_problem_obj} and \eqref{Perfect_problem_const2} involves unit-modulus constraint, which makes it  difficult  to solve. We can  readily prove  that at the optimal solution to \eqref{Perfect_problem}, we must have $P_k^{opt} = {P_{\max }},{k \in {{\cal K}_1}}$ since \eqref{Perfect_problem_obj}  is monotonically increasing  w.r.t. $P_i$. As such,  problem \eqref{Perfect_problem}  can be  simplified to  
\begin{subequations} \label{Perfect_problem_compact}
	\begin{align}
	&\mathop {\max }\limits_{\tau ,{\bf{v}}} \sum\limits_{k = 1}^K {{\tau _k}} {\log _2}\left( {1 + \frac{{{\eta _k}{P_{\max }}\sum\nolimits_{i = 0,i \ne k}^K {{\tau _i}{{\left| {{h_{d,k}} + {\bf{v}}_i^H{{\bf{q}}_k}} \right|}^2}{{\left| {{{\bar h}_{d,k}} + {\bf{v}}_k^H{{{\bf{\bar q}}}_k}} \right|}^2}} }}{{\Gamma {\sigma ^2}{\tau _k}}}} \right)\\
	&{\rm s.t.}~\eqref{Perfect_problem_const3},\eqref{Perfect_problem_const4},\eqref{Perfect_problem_const2}.
	\end{align}
\end{subequations}
Although problem \eqref{Perfect_problem_compact} is still a non-convex optimization problem, it has less constraints and variables compared to  \eqref{Perfect_problem}. Additionally, it is observed that
each optimization variable in \eqref{Perfect_problem_compact} is involved in at most one constraint, which  motivates us to apply AO method to solve it. Specifically, we  divide all the variables into two blocks, i.e., 1) time allocation $\tau$, and  2) phase-shift vector $\bf v$, and then optimize each block in an iterative way, until   convergence is achieved.
\subsection{Time Allocation Optimization}
For any given phase-shift vector $\bf v $, the time allocation optimization  problem is given by  
\begin{subequations} \label{Perfect_subproblem_1}
	\begin{align}
	&\mathop {\max }\limits_\tau  \sum\limits_{k = 1}^K {{\tau _k}} {\log _2}\left( {1 + \frac{{{\eta _k}{P_{\max }}\sum\nolimits_{i = 0,i \ne k}^K {{\tau _i}{{\left| {{h_{d,k}} + {\bf{v}}_i^H{{\bf{q}}_k}} \right|}^2}{{\left| {{{\bar h}_{d,k}} + {\bf{v}}_k^H{{{\bf{\bar q}}}_k}} \right|}^2}} }}{{\Gamma {\sigma ^2}{\tau _k}}}} \right)\\
	&{\rm s.t.}~\eqref{Perfect_problem_const3},\eqref{Perfect_problem_const4}.
	\end{align}
\end{subequations}
\textbf{\emph{Lemma 1:}} The objective function of \eqref{Perfect_subproblem_1} is a jointly concave function of  $\tau$.

\hspace*{\parindent}\textit{Proof}: It can be proved via checking its Hessian matrix, please refer to Appendix~B of reference \cite{kang2015full} for details.

Based on Lemma 1 and the fact that all constraints in \eqref{Perfect_subproblem_1} are linear, problem \eqref{Perfect_subproblem_1} is thus a convex optimization problem, which can be solved by the standard convex optimization techniques, such as interior-point method \cite{boyd2004convex}.

\subsection{IRS Phase Shift Optimization}
 For any given time allocation $\tau$, the IRS phase shift optimization is given by
\begin{subequations} \label{Perfect_subproblem_2}
	\begin{align}
	&\mathop {\max }\limits_{\bf v}  \sum\limits_{k = 1}^K {{\tau _k}} {\log _2}\left( {1 + \frac{{{\eta _k}{P_{\max }}\sum\nolimits_{i = 0,i \ne k}^K {{\tau _i}{{\left| {{h_{d,k}} + {\bf{v}}_i^H{{\bf{q}}_k}} \right|}^2}{{\left| {{{\bar h}_{d,k}} + {\bf{v}}_k^H{{{\bf{\bar q}}}_k}} \right|}^2}} }}{{\Gamma {\sigma ^2}{\tau _k}}}} \right)\\
%	&{\rm s.t.}~\left\| {{{{v}}_{k,m}}} \right\| = 1,\forall m\in{\cal M},{k \in {{\cal K}_1}}. \label{Perfect_subproblem_2_const1}
&{\rm s.t.}~\eqref{Perfect_problem_const2}.
	\end{align}
\end{subequations} 
Problem \eqref{Perfect_subproblem_2} involves    non-convex  unit-modulus constraint   \eqref{Perfect_problem_const2}. To address this non-convexity, 
we   relax the unit-modulus constraint  \eqref{Perfect_problem_const2} as 
\begin{align}
\left\| {{{{v}}_{k,m}}} \right\| \le 1, \forall m\in{\cal M},{k \in {{\cal K}_1}},\label{Perfect_problem_const2_relax}
\end{align}
which is  convex. 

In addition, it is also observed that     ${\bf v}_i,i\ne k,$ does not appear  in  ${\sum\nolimits_{i = 0,i \ne k}^K {{\tau _i}{{\left| {{h_{d,k}} + {\bf{v}}_i^H{{\bf{q}}_k}} \right|}^2}} }$ and ${\left| {{{\bar h}_{d,k}} + {\bf{v}}_k^H{{{\bf{\bar q}}}_k}} \right|^2}$ simultaneously, which motivates us to apply AO algorithm to solve it.  As such,   we can partition the entire phase shifts  into $K+1$ blocks, i.e., $\{{\bf v}_i\},i\in{\cal K}_1$, and then alternately optimize each block until convergence is achieved.  However, optimizing    phase shift ${\bf v}_0$ for DL WET and  ${\bf v}_k,k\in {\cal K}_2$, for  UL WIT are different. As such,  we solve  them  separately with two cases in the following. 

\textit{Case 1}: Define ${y_{k,i}} = {\left| {{h_{d,k}} + {\bf{v}}_i^H{{\bf{q}}_k}} \right|^2}$  and ${{\bar y}_k} = {\left| {{{\bar h}_{d,k}} + {\bf{v}}_k^H{{{\bf{\bar q}}}_k}} \right|^2}$, $k \in {{\cal K}_2},i \in {{\cal K}_1}$. The phase shift optimization problem for DL WET,  i.e., for ${{{\bf{v}}_0}}$,  is given by 
\begin{subequations} \label{perfect_dynamic_phaseshift_wet}
	\begin{align}
	&\mathop {\max }\limits_{{{\bf{v}}_0}} \sum\limits_{k = 1}^K {{\tau _k}} {\log _2}\left( {1 + \frac{{{\eta _k}{P_{\max }}\left( {{\tau _0}{{\left| {{h_{d,k}} + {\bf{v}}_0^H{{\bf{q}}_k}} \right|}^2} + \sum\nolimits_{i = 1,i \ne k}^K {{\tau _i}{y_{k,i}}} } \right){{\bar y}_k}}}{{\Gamma {\sigma ^2}{\tau _k}}}} \right)\label{perfect_dynamic_phaseshift_wet_obj}\\
	&{\rm s.t.}~\left\| {{{{v}}_{0,m}}} \right\| \le 1,\forall m\in{\cal M}. \label{perfect_dynamic_relax_unit_modulus_1}
	\end{align}
\end{subequations} 
It is observed that ${\left| {{h_{d,k}} + {\bf{v}}_0^H{{\bf{q}}_k}} \right|^2}$ is a convex quadratic function of ${{{\bf{v}}_0}}$.  Recall that any  convex function is globally
lower-bounded by its first-order Taylor expansion at any feasible point \cite{boyd2004convex}.  As a result, the SCA technique  is applied. Specifically, for any 
local point ${{{\bf{v}}_i^r}}$ in the $r$th iteration, we have 
\begin{align}
{\left| {{h_{d,k}} + {{\bf{v}}^H_i}{\bf q}_k} \right|^2}& \ge  - \left| {{h_{d,k}} + {\bf{v}}_i^{r,H}{\bf q}_k} \right|^2 + 2{\rm{Re}}\left\{ {{{\left( {{h_{d,k}} + {\bf{v}}_i^H{{\bf{q}}_k}} \right)}^H}\left( {{h_{d,k}} + {\bf{v}}_i^{r,H}{{\bf{q}}_k}} \right)} \right\} \notag\\
&\overset{\triangle}{=} {f_k}\left( {{{\bf{v}}_i}} \right),i \in {{\cal K}_1},k \in {{\cal K}_2}, \label{perfect_dyamic_taylor_1}
\end{align}
which is linear  and convex w.r.t. ${{{\bf{v}}_i}}$. 

For $i=0$, $f_k({\bf v}_0)$ is a lower bound of ${\left| {{h_{d,k}} + {\bf{v}}_0^H{{\bf{q}}_k}} \right|^2}$. Substituting ${f_k}\left( {{{\bf{v}}_0}} \right)$ into \eqref{perfect_dynamic_phaseshift_wet_obj}, we have the following convex optimization  problem 
\begin{subequations} \label{perfect_dynamic_phaseshift_wet_approx}
	\begin{align}
	&\mathop {\max }\limits_{{{\bf{v}}_0}} \sum\limits_{k = 1}^K {{\tau _k}} {\log _2}\left( {1 + \frac{{{\eta _k}{P_{\max }}\left( {{\tau _0}{f_k}\left( {{{\bf{v}}_0}} \right) + \sum\nolimits_{i = 1,i \ne k}^K {{\tau _i}{y_{k,i}}} } \right){{\bar y}_k}}}{{\Gamma {\sigma ^2}{\tau _k}}}} \right)\\
	&{\rm s.t.}~\eqref{perfect_dynamic_relax_unit_modulus_1}, 
	\end{align}
\end{subequations}
which can be solved by the interior point method \cite{boyd2004convex}.

\textit{Case 2}: The phase shift optimization problem for UL WIT,  i.e., for ${{{\bf{v}}_k}}, k\in {\cal K}_2$. We alternately update phase-shift vector over each UL time slot  while others being fixed. The problem is  given by  
\begin{subequations} \label{perfect_dynamic_phaseshift_wit}
	\begin{align}
	&\mathop {\max }\limits_{{{\bf{v}}_k}} {\tau _k}{\log _2}\left( {1 + {\eta _k}{P_{\max }}\sum\nolimits_{i = 0,i \ne k}^K {{\tau _i}{y_{k,i}}} {{\left| {{{\bar h}_{d,k}} + {\bf{v}}_k^H{{{\bf{\bar q}}}_k}} \right|}^2}/\Gamma {\sigma ^2}{\tau _k}} \right) +   \notag \\
	&\sum\limits_{j = 1,j \ne k}^K {{\tau _j}} {\log _2}\left( {1 + {\eta _j}{P_{\max }}\left( {{\tau _k}{{\left| {{h_{d,j}} + {\bf{v}}_k^H{{\bf{q}}_j}} \right|}^2} + \sum\limits_{i = 0,i \ne j,i \ne k}^K {{\tau _i}{y_{j,i}}} } \right){{\bar y}_j}/\Gamma {\sigma ^2}{\tau _j}} \right)\\
	&{\rm s.t.}~\left\| {{{{v}}_{k,m}}} \right\| \le 1,\forall m\in{\cal M}.
	\end{align}
\end{subequations} 
Similar to \eqref{perfect_dyamic_taylor_1}, taking the first-order Taylor expansion of ${\left| {{{\bar h}_{d,k}} + {\bf{v}}_k^H{{{\bf{\bar q}}}_k}} \right|^2}$ at any feasible point ${{{\bf{v}}_k^r}}$,  we have its lower bound given by
\begin{align}
{\left| {{{\bar h}_{d,k}} + {\bf{v}}_k^H{{{\bf{\bar q}}}_k}} \right|^2} &\ge  - {\left| {{{\bar h}_{d,k}} + {\bf{v}}_k^{r,H}{{{\bf{\bar q}}}_k}} \right|^2} + 2{\rm{Re}}\left\{ {{{\left( {{{\bar h}_{d,k}} + {\bf{v}}_k^H{{{\bf{\bar q}}}_k}} \right)}^H}\left( {{{\bar h}_{d,k}} + {\bf{v}}_k^{r,H}{{{\bf{\bar q}}}_k}} \right)} \right\}\notag\\
& \overset{\triangle}{= }{{\bar f}_k}\left( {{{\bf{v}}_k}} \right),k \in {{\cal K}_2}. \label{perfect_dyamic_taylor_2}
\end{align}
It can be readily checked that  ${{\bar f}_k}\left( {{{\bf{v}}_k}} \right)$ is linear  and convex w.r.t. ${{{\bf{v}}_k}}$. Based on \eqref{perfect_dyamic_taylor_1} and \eqref{perfect_dyamic_taylor_2},  problem  \eqref{perfect_dynamic_phaseshift_wit} can be approximated  by 
\begin{subequations} \label{perfect_dynamic_phaseshift_wit_approx}
	\begin{align}
	&\mathop {\max }\limits_{{{\bf{v}}_k}} {\tau _k}{\log _2}\left( {1 + {\eta _k}{P_{\max }}\sum\nolimits_{i = 0,i \ne k}^K {{\tau _i}{y_{k,i}}} {{\bar f}_k}\left( {{{\bf{v}}_k}} \right)/\Gamma {\sigma ^2}{\tau _k}} \right)  +  \notag \\
	&\qquad\sum\limits_{j = 1,j \ne k}^K {{\tau _j}} {\log _2}\left( {1 + {\eta _j}{P_{\max }}\left( {{\tau _k}{f_j}\left( {{{\bf{v}}_k}} \right) + \sum\nolimits_{i = 0,i \ne j,i \ne k}^K {{\tau _i}{y_{j,i}}} } \right){{\bar y}_j}/\Gamma {\sigma ^2}{\tau _j}} \right)\\
	&{\rm s.t.}~\left\| {{{{v}}_{k,m}}} \right\| \le 1,\forall m\in{\cal M},
	\end{align}
\end{subequations}
which is convex and can be solved by interior-point method \cite{boyd2004convex}.
\subsection{Overall Algorithm and Computational  Complexity Analysis}
\begin{algorithm}[!t]
	\caption{AO   for solving  \eqref{Perfect_problem}.}	\label{alg1}
	\begin{algorithmic}[1]
		\STATE  \textbf{Initialize} IRS phase-shift vector  ${{\bf{v}}_k^r}, k \in {{\cal K}_1}$, and  threshold $\varepsilon$.
		\STATE  \textbf{repeat}
		\STATE  \quad Update time allocation by solving  \eqref{Perfect_subproblem_1}.
		\STATE  \quad Update DL WET phase shift  vector ${\bf v}_0$  by solving \eqref{perfect_dynamic_phaseshift_wet_approx}.
		\STATE \quad \textbf{for} $k = 1, \ldots ,K$
		\STATE  \qquad Update UL WIT  phase shift  vector ${\bf v}_k$    by solving \eqref{perfect_dynamic_phaseshift_wit_approx}.
		\STATE \quad \textbf{end}
		\STATE \textbf{until}   the fractional increase of the objective function of \eqref{Perfect_problem} is below a  threshold $\varepsilon$.
		\STATE Reconstruct  phase shifts based on \eqref{phaseshift_reconstruct}.
		\STATE Update time allocation by solving \eqref{Perfect_subproblem_1} based on the newly obtained phase shifts. 
		\STATE {\bf Output:} time allocation ${\tau}$, and phase-shift vector $\bf{v}$.
	\end{algorithmic}
\end{algorithm}
Finally, we need to reconstruct the obtained phase shifts  as  unit-modulus solutions. The reconstruction scheme is given by 
\begin{align}
{v_{k,m}^{opt}} = \frac{{{v_{k,m}}}}{{\left| {{v_{k,m}}} \right|}},k \in {{\cal K}_1},m \in {\cal M}. \label{phaseshift_reconstruct}
\end{align}
Based on the solutions to the above subproblems, an AO  algorithm is proposed, which is   summarized in Algorithm~\ref{alg1}.  

The mainly computational complexity lies from   steps $3$,  $4$, and   $6$. Specifically, in  step $3$,  \eqref{Perfect_subproblem_1} can be solved by the interior-point method, whose complexity is ${\cal O}\left( {K{\rm{ + }}1} \right)^{3.5}$ \cite{gondzio1996computational}, where $K+1$ denotes the number of variables. In  steps $4$ and $6$, the complexity for solving \eqref{perfect_dynamic_phaseshift_wet_approx} and \eqref{perfect_dynamic_phaseshift_wit_approx} by the interior-point method is the  same with 
${\cal O}{\left( M^{3.5} \right)}$, where  $M$ denotes the  number of variables. Therefore, the total complexity of Algorithm~\ref{alg1} is ${\cal O}\left( {{L_{iter}}\left( {\left( {K{\rm{ + }}1} \right){M^{3.5}} + {{\left( {K{\rm{ + }}1} \right)}^{3.5}}} \right)} \right)$, where ${{L_{iter}}}$ stands for the number of iterations required to reach convergence.  Since  at   steps $3$,  $4$, and  $6$, each subproblem is optimally solved,  the objective function  is   non-decreasing over iterations. In addition, the maximum  objective value of \eqref{Perfect_problem} is upper bounded by a finite value due
to the limited HAP transmit power. As such, by
applying the proposed Algorithm~\ref{alg1}, the objective value is guaranteed to be non-decreasing over the
iterations and terminated finally.
\section{Proposed Solution for  Fully  Dynamic IRS Beamforming    with Imperfect SIC}
For the fully  dynamic IRS beamforming     with imperfect SIC, i.e., $\gamma \ne0$, the formulated problem is more challenging than  that with  perfect SIC  since the HAP transmission  power $P$ is involved in the objective function of \eqref{p1}. To solve this problem, we extend the proposed AO algorithm    in Section III to a novel  penalty-based algorithm, which includes a two-layer iteration, i.e., an inner layer iteration and an outer layer iteration. The inner layer solves
the penalized optimization problem by exploiting the AO algorithm, while the outer layer updates the penalty coefficient over iterations to guarantee convergence. Specifically, 
by introducing a new  auxiliary variable $z=\left\{ {{z_k},k \in {{\cal K}_2}} \right\}$,  problem \eqref{p1} is rewritten as 
\begin{subequations} \label{p1_1}
	\begin{align}
	&\mathop {\max }\limits_{\tau ,P,{\bf{v}},z} \sum\limits_{k = 1}^K {{\tau _k}} {\log _2}\left( {1 + \frac{{{\eta _k}\sum\nolimits_{i = 0,i \ne k}^K {{P_i}} {\tau _i}{{\left| {{h_{d,k}} + {\bf{v}}_i^H{{\bf{q}}_k}} \right|}^2}{{\left| {{{\bar h}_{d,k}} + {\bf{v}}_k^H{{{\bf{\bar q}}}_k}} \right|}^2}}}{{{\sigma ^2}{z_k}{\tau _k}}}} \right)\\
	&{\rm s.t.}~{z_k} = \Gamma \left( {\beta \gamma {P_k}/{\sigma ^2} + 1} \right),k \in {{\cal K}_2},\label{p1_1_const1}\\
	&\qquad \eqref{Perfect_problem_const3},\eqref{Perfect_problem_const4},\eqref{Perfect_problem_const5},\eqref{Perfect_problem_const2}.
	\end{align}
\end{subequations}
We then use \eqref{p1_1_const1} as a penalty term that is added to the objective function of \eqref{p1_1}, yielding  the following optimization problem 
\begin{subequations} \label{p1_1_penalty}
	\begin{align}
	&\mathop {\max }\limits_{\tau ,P,{\bf{v}},z} \sum\limits_{k = 1}^K {{\tau _k}} {\log _2}\left( {1 + \frac{{{\eta _k}\sum\nolimits_{i = 0,i \ne k}^K {{P_i}} {\tau _i}{{\left| {{h_{d,k}} + {\bf{v}}_i^H{{\bf{q}}_k}} \right|}^2}{{\left| {{{\bar h}_{d,k}} + {\bf{v}}_k^H{{{\bf{\bar q}}}_k}} \right|}^2}}}{{{\sigma ^2}{z_k}{\tau _k}}}} \right)\notag\\
	& \qquad- \frac{1}{{2\rho }}\sum\limits_{k = 1}^K {{{\left| {{z_k} - \Gamma \left( {\beta \gamma {P_k}/{\sigma ^2} + 1} \right)} \right|}^2}} \\
	&{\rm s.t.} ~\eqref{Perfect_problem_const3},\eqref{Perfect_problem_const4},\eqref{Perfect_problem_const5},\eqref{Perfect_problem_const2},
	\end{align}
\end{subequations}
where  $\rho> 0$ is a penalty coefficient that penalizes the violation equality constraint \eqref{p1_1_const1}. By gradually decreasing the value of $\rho$  in the outer layer,   as $\rho  \to 0$, it follows  that $\frac{1}{{2\rho }} \to \infty $. In this case, equality in \eqref{p1_1_const1} is guaranteed in the optimal solution to problem \eqref{p1_1_penalty}. For any given $\rho>0$, \eqref{p1_1} is still a non-convex optimization problem due to the non-convex
objective function as well as non-convex unit-modulus  constraint  \eqref{Perfect_problem_const2}. However, it is observed that each optimization variable is involved in at most one constraint, which motivates us to apply AO algorithm  to solve it in the inner layer. Specifically, we first relax unit-modulus constraint \eqref{Perfect_problem_const2} as  \eqref{Perfect_problem_const2_relax}. We then divide all the optimization variables into four blocks, i.e., 1) time allocation $\tau$, 2) HAP transmission power $P$, 3) auxiliary variable $z$, and 4) IRS phase shift $\bf v$, and then alternately optimize each block, until   convergence is achieved.
\subsubsection{Optimizing $\tau$ for given ${P,{\bf{v }},z}$} Ignoring irrelevant terms w.r.t. $\tau$, this subproblem can be expressed as
\begin{subequations}\label{p1_1_penalty_inner1}
\begin{align}
&\mathop {\max }\limits_{\tau} \sum\limits_{k = 1}^K {{\tau _k}} {\log _2}\left( {1 + \frac{{{\eta _k}\sum\nolimits_{i = 0,i \ne k}^K {{P_i}} {\tau _i}{{\left| {{h_{d,k}} + {\bf{v}}_i^H{{\bf{q}}_k}} \right|}^2}{{\left| {{{\bar h}_{d,k}} + {\bf{v}}_k^H{{{\bf{\bar q}}}_k}} \right|}^2}}}{{{\sigma ^2}{z_k}{\tau _k}}}} \right)\\
	&{\rm s.t.}~\eqref{Perfect_problem_const3},\eqref{Perfect_problem_const4}.
\end{align}
\end{subequations} 
Problem \eqref{p1_1_penalty_inner1} has a similar form to problem \eqref{Perfect_subproblem_1} discussed in  Section III-A,  which thus can be solved similarly.
\subsubsection{Optimizing  $P$  for given $\tau,{\bf{v }},z$}This subproblem is written as
\begin{subequations} \label{p1_1_penalty_inner2}
\begin{align}
&\mathop {\max }\limits_{P} \sum\limits_{k = 1}^K {{\tau _k}} {\log _2}\left( {1 + \frac{{{\eta _k}\sum\nolimits_{i = 0,i \ne k}^K {{P_i}} {\tau _i}{{\left| {{h_{d,k}} + {\bf{v}}_i^H{{\bf{q}}_k}} \right|}^2}{{\left| {{{\bar h}_{d,k}} + {\bf{v}}_k^H{{{\bf{\bar q}}}_k}} \right|}^2}}}{{{\sigma ^2}{z_k}{\tau _k}}}} \right)\notag\\
& \qquad- \frac{1}{{2\rho }}\sum\limits_{k = 1}^K {{{\left| {{z_k} - \Gamma \left( {\beta \gamma {P_k}/{\sigma ^2} + 1} \right)} \right|}^2}} \\
&{\rm s.t.}~\eqref{Perfect_problem_const5}.
\end{align}
\end{subequations} 
It is observed that both the objective function and  constraints  are convex, which  thus  can be efficiently solved by  interior-point method \cite{boyd2004convex}.
\subsubsection{Optimizing  $z$  for given $P,\tau,{\bf{v }}$}This subproblem can be expressed as
\begin{align}\label{p1_1_penalty_inner3}
&\mathop {\max }\limits_{z} \sum\limits_{k = 1}^K {{\tau _k}} {\log _2}\left( {1 + \frac{{{\eta _k}\sum\nolimits_{i = 0,i \ne k}^K {{P_i}} {\tau _i}{{\left| {{h_{d,k}} + {\bf{v}}_i^H{{\bf{q}}_k}} \right|}^2}{{\left| {{{\bar h}_{d,k}} + {\bf{v}}_k^H{{{\bf{\bar q}}}_k}} \right|}^2}}}{{{\sigma ^2}{z_k}{\tau _k}}}} \right)\notag\\
& \qquad- \frac{1}{{2\rho }}\sum\limits_{k = 1}^K {{{\left| {{z_k} - \Gamma \left( {\beta \gamma {P_k}/{\sigma ^2} + 1} \right)} \right|}^2}}
\end{align}
It is observed that \eqref{p1_1_penalty_inner3}  has no constraints and the  auxiliary optimization variables  $\{z_k, k\in{\cal K}_2\}$ can be solved  in parallel.  Without loss of generality,  the $k$th subproblem of \eqref{p1_1_penalty_inner3} is given by
\begin{align}
\mathop {\max }\limits_{{z_k}} {\tau _k}{\log _2}\left( {1 + \frac{{{a_k}}}{{{z_k}}}} \right) - \frac{1}{{2\rho }}{\left| {{z_k} - \Gamma \left( {\beta \gamma {P_k}/{\sigma ^2} + 1} \right)} \right|^2},\label{p1_1_penalty_inner3_kth}
\end{align}
where ${a_k} = {\eta _k}\sum\nolimits_{i = 0,i\ne k}^{k - 1} {{\tau _i}{P_i}{{\left| {{h_{d,k}} + {\bf{v}}_i^H{{\bf{q}}_k}} \right|}^2}} \left| {{{\bar h}_{d,k}} + {\bf{v}}_k^H{{{\bf{\bar q}}}_k}} \right|/{\tau _k}{\sigma ^2},k \in {{\cal K}_2}$.
The key observation in \eqref{p1_1_penalty_inner3_kth} is that the function ${\log _2}\left( {1 + \frac{{{a_k}}}{{{z_k}}}} \right)$ is convex w.r.t. ${{z_k}}$, and its lower bound can be obtained by taking its first-order Taylor expansion at any feasible point. As a result,  the SCA technique   is applied here. Specifically,  for any given point ${z_k^r}$ in  the $r$th iteration,  we have 
\begin{align}
{\log _2}\left( {1 + \frac{{{a_k}}}{{{z_k}}}} \right) \ge {\log _2}\left( {1 + \frac{{{a_k}}}{{z_k^r}}} \right) - \frac{{{a_k}{{\log }_2}e}}{{z_k^r\left( {z_k^r + {a_k}} \right)}}\left( {{z_k} - z_k^r} \right). \label{p1_1_penalty_inner3_const1}
\end{align}
Replacing  ${\log _2}\left( {1 + \frac{{{a_k}}}{{{z_k}}}} \right)$ by its lower bound, we have the newly formulated optimization problem  as follows
\begin{align}\label{imperfect_SI_SCA}
\mathop {\max }\limits_{{z_k}} {\tau _k}\left( {{{\log }_2}\left( {1 + \frac{{{a_k}}}{{z_k^r}}} \right) - \frac{{{a_k}{{\log }_2}e}}{{z_k^r\left( {z_k^r + {a_k}} \right)}}\left( {{z_k} - z_k^r} \right)} \right) - \frac{1}{{2\rho }}{\left| {{z_k} - \Gamma \left( {\frac{{\beta \gamma {P_k}}}{{{\sigma ^2}}} + 1} \right)} \right|^2}
\end{align}
which is convex.  By taking the  first-order derivative of    \eqref{imperfect_SI_SCA} w.r.t. ${{z_k}}$  and setting it to zero, the closed-form optimal solution to  \eqref{imperfect_SI_SCA} can be obtained as 
\begin{align}
{z_k^{opt}} =  \Gamma \left( {\frac{{\beta \gamma {P_k}}}{{{\sigma ^2}}} + 1} \right)- \frac{{\rho {a_k}{\tau _k}{{\log }_2}e}}{{z_k^r\left( {z_k^r + {a_k}} \right)}},k \in {\cal K}_2. \label{p1_auxitliy_optimal}
\end{align}

\subsubsection{Optimizing  ${\bf{v }}$  for given $P,\tau,z$}  Ignoring irrelevant terms w.r.t. ${\bf{v }}$, this subproblem can be expressed as  
\begin{subequations}\label{p1_1_penalty_inner4}
\begin{align}
&\mathop {\max }\limits_{\bf v} \sum\limits_{k = 1}^K {{\tau _k}} {\log _2}\left( {1 + \frac{{{\eta _k}\sum\nolimits_{i = 0,i \ne k}^K {{P_i}} {\tau _i}{{\left| {{h_{d,k}} + {\bf{v}}_i^H{{\bf{q}}_k}} \right|}^2}{{\left| {{{\bar h}_{d,k}} + {\bf{v}}_k^H{{{\bf{\bar q}}}_k}} \right|}^2}}}{{{\sigma ^2}{z_k}{\tau _k}}}} \right)\\
&{\rm s.t.} ~\eqref{Perfect_problem_const2_relax}.
\end{align}
\end{subequations}
Since  problem \eqref{p1_1_penalty_inner4} has a similar form to \eqref{Perfect_subproblem_2} discussed in Section III-B,  which can be solved with the same way and   is omitted here for brevity.
\subsubsection{Overall Algorithm and Computational  Complexity Analysis}
In the outer layer, we gradually decrease the value of penalty coefficient $\rho^r$ in the $r$th iteration as follows 
\begin{align}
{\rho ^r} = c{\rho ^{r - 1}}, 0\le c\le 1, \label{update_penalty_coefficient}
\end{align}
where $c$ is a  step size. Generally, a larger value of $c$ can achieve better performance but at the cost of more iterations in the outer layer. It is  worth pointing out that if $c$ is chosen very small, the penalty-based algorithm will  be eventually diverged with a high possibility. Empirically, choosing $c$  from $0.7$ to $0.9$ is a good choice to  balance the convergence speed and performance.
To evaluate whether it violates the equality  constraint 
\eqref{p1_1_const1} or not, we adopt an indicator $\xi$ defined as
\begin{align}
\xi  = \mathop {\max }\limits_{k \in {{\cal K}_2}} \left\{ {\left| {{z_k} - \Gamma \left( {\beta \gamma {P_k}/{\sigma ^2} + 1} \right)} \right|} \right\}.
\end{align}
The algorithm is terminated when $\xi$  is  below a predefined threshold.
\begin{algorithm}[!t]
	\caption{Penalty-based algorithm for solving problem \eqref{p1}.}
	\label{alg2}
	\begin{algorithmic}[1]
		\STATE  \textbf{Initialize} ${\bf v}^r$, $P^r$, $z^r$, $c$, $\rho$, $\varepsilon_1$, and  $\varepsilon_2$.
		\STATE  \textbf{repeat: outer layer}
		\STATE \quad \textbf{repeat: inner layer }
		\STATE  \qquad Update $\tau$ by solving  \eqref{p1_1_penalty_inner1}.
		\STATE  \qquad Update $P$ by solving \eqref{p1_1_penalty_inner2}.
		\STATE  \qquad Update $z$ based on \eqref{p1_auxitliy_optimal}.
		\STATE  \qquad Update ${\bf{v }}$ by solving  \eqref{p1_1_penalty_inner4}.
		\STATE \quad \textbf{until} the fractional increase of the objective value of \eqref{p1_1_penalty} is below a threshold $\varepsilon_1$.
		\STATE  \quad Update penalty coefficient $\rho$ based on \eqref{update_penalty_coefficient}.
		\STATE \textbf{until}  penalty  violation $\xi$ is below a threshold $\varepsilon_2$.
		\STATE Reconstruct  phase shifts based on \eqref{phaseshift_reconstruct}.
\STATE Update time allocation $\tau$, power allocation $P$, and auxiliary variable $z$  by solving \eqref{p1_1_penalty_inner1}, \eqref{p1_1_penalty_inner2}, and \eqref{p1_auxitliy_optimal}, respectively,  based on the new phase shifts. 
\STATE {\bf Output:} time allocation ${\tau}$, power allocation $P$, and phase shift $\bf{v}$.
	\end{algorithmic}
\end{algorithm}
 The details of this algorithm are summarized in Algorithm~\ref{alg2}.   Note that with the proper variable partition in our proposed algorithm, there is no  constraint coupling between the variables in different blocks, as seen in \eqref{p1_1_penalty_inner1}, \eqref{p1_1_penalty_inner2}, \eqref{p1_auxitliy_optimal}, and \eqref{p1_1_penalty_inner4}. In addition, each subproblem is either  locally or optimally   solved.
As such, based on the results in \cite{shi2016joint}, the obtained solution converges to a point fulfilling the Karush–Kuhn–Tucker (KKT) optimality conditions of original problem \eqref{p1}.
 
 The complexity of Algorithm~\ref{alg2} can be analyzed  as follows. In the inner layer, the main complexity of Algorithm~\ref{alg2} comes from steps $4$, $5$ and $7$,  which can be calculated similarly as Algorithm~\ref{alg1} discussed in Section III-C. Therefore, the total complexity of Algorithm~\ref{alg2} is ${\cal O}\left( {{L_{outer}}{L_{inner}}\left( {2{{\left( {K{\rm{ + }}1} \right)}^{3.5}} + \left( {K{\rm{ + }}1} \right){M^{3.5}}} \right)} \right)$, where ${{L_{inner}}}$ and ${{L_{outer}}}$
 denote the numbers of iterations required for reaching convergence in the inner layer and outer layer,  respectively.
  \section{Extension to Partially Dynamic  and Static IRS Beamforming}
  In this section, we study two special cases, namely, partially dynamic IRS beamforming and static IRS beamforming,  for both perfect SIC and imperfect SIC. Since  the  problem formulations  of  the  above two special cases  are different from that of  fully  dynamic IRS beamforming in \eqref{p1}, the new algorithms are   required. However, it can be easily observed that the resource allocation (transmit power $P$ and time allocation $\tau$) optimization of problems \eqref{Partially_p} and \eqref{pStatic} can be solved  in  the same way as \eqref{p1} for both perfect SIC and imperfect case discussed in Section III and Section IV, respectively. As such, in this section, we only focus on the  phase shift optimization for the  partially dynamic IRS beamforming and static IRS beamforming as below. Without loss of generality, we first solve the phase shift optimization for the static IRS case. Then, we will show later that the phase shift optimization for the partially dynamic beamforming can be solved by a combination of that for fully dynamic beamforming  and static beamforming.
 \subsection{Phase Shift Optimization for   Static IRS Beamforming with Imperfect SIC}
 For the static IRS beamforming with the  imperfect SIC, problem \eqref{pStatic} is more challenging than problem \eqref{p1} since phase-shift vector $\bf v$ appears   in both ${\left| {{h_{d,k}} + {\bf{v}}_s^H{{\bf{q}}_k}} \right|^2}$ and ${\left| {{{\bar h}_{d,k}} + {\bf{v}}_s^H{{{\bf{\bar q}}}_k}} \right|^2}$. As a result,  the proposed SCA  for the phase shift optimization in \eqref{p1}  cannot be applied to  \eqref{pStatic}. To tackle this difficulty, a DC optimization
 method is proposed with at least a locally optimal solution guaranteed. Specifically,    the phase shift optimization subproblem for  \eqref{pStatic} is given by
  \begin{subequations} \label{pstatic_phaseshift}
 	\begin{align}
 	&\mathop {\max }\limits_{{{\bf{v}}_s}} \sum\limits_{k = 1}^K {{\tau _k}} {\log _2}\left( {1 + \frac{{{\eta _k}\left( {\sum\nolimits_{i = 0,i \ne k}^K {{P_i}} {\tau _i}} \right){{\left| {{h_{d,k}} + {\bf{v}}_s^H{{\bf{q}}_k}} \right|}^2}{{\left| {{{\bar h}_{d,k}} + {\bf{v}}_s^H{{{\bf{\bar q}}}_k}} \right|}^2}}}{{\Gamma \left( {\beta \gamma {P_k} + {\sigma ^2}} \right){\tau _k}}}} \right)\\
 	&{\rm s.t.} ~\eqref{pstatic_const_1}.
 	\end{align}
 \end{subequations}
Let   
${\left| {{h_{d,k}} + {\bf{v}}_s^H{{\bf{q}}_k}} \right|^2} = {\left| {{{{\bf{\tilde v}}}^H}{{\bf{h}}_k}} \right|^2}$ and ${\left| {{{\bar h}_{d,k}} + {\bf{v}}_s^H{{{\bf{\bar q}}}_k}} \right|^2} = {\left| {{{{\bf{\tilde v}}}^H}{{{\bf{\bar h}}}_k}} \right|^2}$, where ${{{\bf{\tilde v}}}^H} = \left[ {1{\kern 1pt} {\kern 1pt} {\kern 1pt} {\kern 1pt} {\kern 1pt} {\kern 1pt} {\kern 1pt} {\kern 1pt} {{\bf{v}}_s^H}} \right]$, ${\bf{h}}_k^H = \left[ {h_{d,k}^H{\kern 1pt} {\kern 1pt} {\kern 1pt} {\kern 1pt} {\kern 1pt} {\kern 1pt} {\kern 1pt} {\kern 1pt} {\kern 1pt} {\kern 1pt} {\kern 1pt} {\kern 1pt} {\kern 1pt} {\bf{q}}_k^H} \right]$ and ${\bf{\bar h}}_k^H = \left[ {\bar h_{d,k}^H{\kern 1pt} {\kern 1pt} {\kern 1pt} {\kern 1pt} {\kern 1pt} {\kern 1pt} {\kern 1pt} {\kern 1pt} {\kern 1pt} {\kern 1pt} {\kern 1pt} {\kern 1pt} {\kern 1pt} {\bf{\bar q}}_k^H} \right]$.
Define ${{\bf{H}}_k} = {{\bf{h}}_k}{\bf{h}}_k^H$, ${{{\bf{\bar H}}}_k} = {{{\bf{\bar h}}}_k}{\bf{\bar h}}_k^H$, ${\bf{\tilde V}} = {\bf{\tilde v}}{{{\bf{\tilde v}}}^H}$, which needs to satisfy ${\bf{\tilde V}} \succeq {\bf{0}}$ and ${\rm{rank}}\left( {{\bf{\tilde V}}} \right) = 1$.  
We thus have 
\begin{align}
{\left| {{h_{d,k}} + {\bf{v}}_s^H{{\bf{q}}_k}} \right|^2}{\left| {{{\bar h}_{d,k}} + {\bf{v}}_s^H{{{\bf{\bar q}}}_k}} \right|^2} = {\left| {{{{\bf{\tilde v}}}^H}{{\bf{h}}_k}} \right|^2}{\left| {{{{\bf{\tilde v}}}^H}{{{\bf{\bar h}}}_k}} \right|^2} = {\rm{tr}}\left( {{\bf{\tilde V}}{{{\bf{\bar H}}}_k}{\bf{\tilde V}}{{\bf{H}}_k}} \right).
\end{align}
As such, we can rewritten \eqref{pstatic_phaseshift} as 
  \begin{subequations} \label{pstatic_phaseshift_SDP}
	\begin{align}
	&\mathop {\max }\limits_{{\bf{\tilde V}} \succeq {\bf 0}} \sum\limits_{k = 1}^K {{\tau _k}} {\log _2}\left( {1 + \frac{{{\eta _k}\left( {\sum\nolimits_{i = 0,i \ne k}^K {{P_i}} {\tau _i}} \right){\rm{tr}}\left( {{\bf{\tilde V}}{{{\bf{\bar H}}}_k}{\bf{\tilde V}}{{\bf{H}}_k}} \right)}}{{\Gamma \left( {\beta \gamma {P_k} + {\sigma ^2}} \right){\tau _k}}}} \right)\label{pstatic_phaseshift_SDP_obj}\\
	&{\rm s.t.} ~{{{\bf{\tilde V}}}_{m,m}} = 1,m = 1, \ldots ,M + 1,\label{pstatic_phaseshift_SDP_const1}\\
	&\qquad {\rm{rank}}\left( {{\bf{\tilde V}}} \right) = 1. \label{pstatic_phaseshift_SDP_const2}
	\end{align}
\end{subequations}
Problem \eqref{pstatic_phaseshift_SDP} is still non-convex due to the non-convexity of the  objective function \eqref{pstatic_phaseshift_SDP_obj}  and  rank-one constraint \eqref{pstatic_phaseshift_SDP_const2}. The key observation in \eqref{pstatic_phaseshift_SDP} is that ${{\rm{tr}}\left( {{\bf{\tilde V}}{{{\bf{\bar H}}}_k}{\bf{\tilde V}}{{\bf{H}}_k}} \right)}$ is convex w.r.t. ${{\bf{\tilde V}}}$, which  motivates us to apply SCA  technique to linearize it into a linear form. Specifically, for any given point ${\bf{\tilde V}}^r$, we have  the following lower bound
\begin{align}
{\rm{tr}}\left( {{\bf{\tilde V}}{{{\bf{\bar H}}}_k}{\bf{\tilde V}}{{\bf{H}}_k}} \right) \ge {\rm{tr}}\left( {{{{\bf{\tilde V}}}^r}{{{\bf{\bar H}}}_k}{{{\bf{\tilde V}}}^r}{{\bf{H}}_k}} \right) + 2{\rm tr}\left( {{{{\bf{\bar H}}}_k}{{{\bf{\tilde V}}}^r}{{\bf{H}}_k}\left( {{\bf{\tilde V}} - {{{\bf{\tilde V}}}^r}} \right)} \right)\overset{\triangle} {=} \tilde f_k\left( {{\bf{\tilde V}}} \right), k\in{\cal K}_2, \label{pstatic_phaseshift_SDP_taylorexpansion}
\end{align}
where the right-hand-side  of \eqref{pstatic_phaseshift_SDP_taylorexpansion} is linear w.r.t $\bf{ \tilde V}$.  

With \eqref{pstatic_phaseshift_SDP_taylorexpansion}, problem \eqref{pstatic_phaseshift_SDP} is approximated as
   \begin{subequations} \label{pstatic_phaseshift_SDP_approx}
 	\begin{align}
 	&\mathop {\max }\limits_{{\bf{\tilde V}} \succeq {\bf 0}} \sum\limits_{k = 1}^K {{\tau _k}} {\log _2}\left( {1 + \frac{{{\eta _k}\left( {\sum\nolimits_{i = 0,i \ne k}^K {{P_i}} {\tau _i}} \right){{\tilde f}_k}\left( {{\bf{\tilde V}}} \right)}}{{\Gamma \left( {\beta \gamma {P_k} + {\sigma ^2}} \right){\tau _k}}}} \right)\label{pstatic_phaseshift_SDP_approx_obj}\\
 	&{\rm s.t.} ~\eqref{pstatic_phaseshift_SDP_const1}, \eqref{pstatic_phaseshift_SDP_const2}.
 	\end{align}
 \end{subequations}
 Note that by removing the rank-one constraint \eqref{pstatic_phaseshift_SDP_const2}, problem  \eqref{pstatic_phaseshift_SDP_approx} becomes a convex semidefinite programming (SDP) problem, which can be efficiently solved by the standard convex techniques. However, the obtained   ${{\bf{\tilde V}}}$ may not be  a rank-one solution. Although the Gaussian randomization technique can be applied to construct a rank-one
 solution from the obtained high-rank solution ${{\bf{\tilde V}}}$, it may not be able to guarantee a locally and/or globally optimal solution, especially
 when the dimension of matrix ${{\bf{\tilde V}}}$ is large  \cite{sidiropoulos2006transmit}. To address this issue,  we apply the DC optimization method to solve \eqref{pstatic_phaseshift_SDP_approx}, which guarantees to converge to a KKT point \cite{tao1997convex}. The main idea behind of DC optimization method  lies in the following equivalence
 \begin{align}
 {\rm{rank}}\left( {{\bf{\tilde V}}} \right) = 1 \Leftrightarrow {\rm{tr}}\left( {{\bf{\tilde V}}} \right) - {\left\| {{\bf{\tilde V}}} \right\|_2} = 0,
 \end{align}
 which indicates that the rank-one constraint  can be equivalently replaced by a DC framework.  Adding  ${\rm{tr}}\left( {{\bf{\tilde V}}} \right) - {\left\| {{\bf{\tilde V}}} \right\|_2}$  as the penalty term in the  objective function \eqref{pstatic_phaseshift_SDP_approx_obj}, problem \eqref{pstatic_phaseshift_SDP_approx}  can be transformed as 
    \begin{subequations} \label{pstatic_phaseshift_SDP_approx_DC}
 	\begin{align}
 	&\mathop {\max }\limits_{{\bf{\tilde V}}\succeq{\bf 0}} \sum\limits_{k = 1}^K {{\tau _k}} {\log _2}\left( {1 + \frac{{{\eta _k}\left( {\sum\nolimits_{i = 0,i \ne k}^K {{P_i}} {\tau _i}} \right){{\tilde f}_k}\left( {{\bf{\tilde V}}} \right)}}{{\Gamma \left( {\beta \gamma {P_k} + {\sigma ^2}} \right){\tau _k}}}} \right) - \frac{1}{{2\rho }}\left( {{\rm{tr}}\left( {{\bf{\tilde V}}} \right) - {{\left\| {{\bf{\tilde V}}} \right\|}_2}} \right)\\
 	&{\rm s.t.} ~\eqref{pstatic_phaseshift_SDP_const1},
 	\end{align}
 \end{subequations}
where $\rho$ is a  penalty coefficient that is  defined in \eqref{p1_1_penalty}.
 Since  ${{{\left\| {{\bf{\tilde V}}} \right\|}_2}}$ is convex w.r.t. ${{\bf{\tilde V}}}$,   we can apply SCA to linearize ${\left\| {{\bf{\tilde V}}} \right\|_2}$. According to \cite{Yang2020Federated}, for any given point ${{{{\bf{\tilde V}}}^r}}$ in the $r$th iteration, by  ignoring irrelevant terms w.r.t. ${{{{\bf{\tilde V}}}}}$, we have the following lower bound 
    \begin{subequations} \label{pstatic_phaseshift_SDP_approx_DC_lowerbound}
 	\begin{align}
 	&\mathop {\max }\limits_{{\bf{\tilde V}}\succeq{\bf{0}}} \sum\limits_{k = 1}^K {{\tau _k}} {\log _2}\left( {1 + \frac{{{\eta _k}\left( {\sum\nolimits_{i = 0,i \ne k}^K {{P_i}} {\tau _i}} \right){{\tilde f}_k}\left( {{\bf{\tilde V}}} \right)}}{{\Gamma \left( {\beta \gamma {P_k} + {\sigma ^2}} \right){\tau _k}}}} \right) -\notag\\
 	&\qquad  \frac{1}{{2\rho }}{\rm{tr}}\left( {\left( {{\bf{I}} - {\bm \lambda} \left( {{{{\bf{\tilde V}}}^r}} \right){{\bm \lambda} ^H}\left( {{{{\bf{\tilde V}}}^r}} \right)} \right){\bf{\tilde V}}} \right)\\
 	&{\rm s.t.} ~\eqref{pstatic_phaseshift_SDP_const1},
 	\end{align}
 \end{subequations}
 where ${{{\bm{\lambda }}^H}\left( {{{{\bf{\tilde V}}}^r}} \right)}$ denotes the eigenvector corresponding to the largest eigenvalue of ${{{{\bf{\tilde V}}}^r}}$. It can be seen that
 both the objective function and constraints  are  convex, which thus can be efficiently solved by the standard convex optimization techniques \cite{boyd2004convex}. By gradually decreasing $\rho$ and successively updating ${{\bf{\tilde V}}}$ by solving \eqref{pstatic_phaseshift_SDP_approx_DC_lowerbound}, the convergence will be finally reached to obtain  at least a locally optimal solution \cite{tao1997convex}, \cite{Yang2020Federated}. 
 
\textbf{\emph{Remark 1:}}   Note that the above proposed DC optimization  is  irrespective of the HAP  transmit power $P$ when it is   fixed,  it is thus 
 applicable to the  static IRS beamforming with perfect SIC. Therefore, we can solve problem \eqref{pStatic} for perfect SIC similarly as in Algorithm~\ref{alg1} and for  imperfect SIC similarly as in Algorithm~\ref{alg2}, with  slight modifications, which are omitted here for brevity.
  \subsection{Phase Shift Optimization for    Partially Dynamic  IRS Beamforming}
 It can be observed  that  phase-shift vectors ${{{\bf{v}}_d}}$ and ${{{\bf{v}}_u}}$ in \eqref{Partially_p} are intricately   coupled, which motivates us to alternately optimize each phase-shift vector. 
 For the fixed ${{{\bf{v}}_d}}$, it can be seen in the objective function of \eqref{Partially_p},  two quadratic functions (i.e., ${{{\left| {{h_{d,k}} + {\bf{v}}_u^H{{\bf{q}}_k}} \right|}^2}}$ and ${{{\left| {{{\bar h}_{d,k}} + {\bf{v}}_u^H{{{\bf{\bar q}}}_k}} \right|}^2}}$) w.r.t. ${{{\bf{v}}_u}}$ are  multiplied together.  As such, it  can also be solved by the DC optimization method   proposed in Section V-A. For the fixed ${{{\bf{v}}_u}}$, there exists one quadratic function (i.e., ${{{\left| {{h_{d,k}} + {\bf{v}}_d^H{{\bf{q}}_k}} \right|}^2}}$) w.r.t. ${{{\bf{v}}_d}}$, which can be solved by the SCA technique proposed in Section III-B.

 \textbf{\emph{Remark 2:}} Similarly, the combination of DC optimization and SCA techniques for phase shift optimization   is also applicable to the partially dynamic  IRS beamforming with perfect SIC. Therefore, we can solve \eqref{pStatic} for perfect SIC similarly as in Algorithm~\ref{alg1} and for  imperfect SIC similarly as in Algorithm~\ref{alg2}, with  slight modifications, which are omitted here for brevity.
 \section{Numerical Results}
 In this section, we provide numerical results   to demonstrate the effectiveness of the proposed algorithms  and to provide useful insights for the IRS-aided FD-WPCN. We consider a system that operates on a carrier frequency of $750~{\rm MHz}$ with the system bandwidth of $1~{\rm MHz}$ and   effective noise power density   $-150~{\rm  dBm/Hz}$ \cite{wu2020jointActive}. We assume that the  IRS  is equipped with  a uniform rectangular array with $M=M_xM_z$ reflecting elements, where $M_x$ and $M_z$ denotes the numbers of reflecting elements along the $x$-axis and $z$-axis, respectively. We fix $M_x=5$ and increase $M_z$ linearly with $M$.  We assume that the antenna spacing is half of the wavelength, i.e., $\lambda /2 = 0.2~\rm m$.  A three dimensional   coordinate setup is considered, where the HAP and the IRS are located at $( 0, 0,0)$, $(10 ~\rm m, 0, 2.5 ~\rm m)$ measured in meter (m), while the devices are uniformly and randomly distributed in a circle centered at $\left( {10~{\rm{m}},0,0 } \right)$ with a radius $1.5~\rm m$. The distance-dependent path loss model is given by $L\left( d \right) = {c_0}{\left( {d/{d_0}} \right)^{ - \alpha }}$,
where ${c_0} = {\left( {\lambda /(4\pi) } \right)^2}$ is the path loss at the reference distance $d_0=1$ m,  $d$ is the link distance, and $\alpha$ is the path loss exponent. 
 We assume that  the HAP-IRS link,  the IRS-device link, and  the HAP-device link    follow Rician fading with a Rician factor of $3~\rm dB$. In addition, the path loss exponents for the HAP-IRS link,  the IRS-device link, and  the HAP-device link are set as $2.2$, $2.2$, and $2.6$, respectively.  Unless otherwise specified, we set $T=1~s$, $\beta=-60~ {\rm dB}$, $\Gamma=9.8~ {\rm dB}$, $\rho=100$, $c=0.85$, $\varepsilon=\varepsilon_1=10^{-2}$, and $\varepsilon_2=10^{-5}$.
%\subsection{Convergence of Algorithm~\ref{alg2}}
%\begin{figure}[!t]
%	\centerline{\includegraphics[width=3.5in]{convergence.eps}}
%	\caption{A random snapshot for showing convergence behaviour of Algorithm~\ref{alg2}.} \label{fig_convergence}
%\end{figure}

\begin{figure}[!t]
	\centering
		\begin{minipage}[t]{0.49\textwidth}
		\centering
		\includegraphics[width=3in]{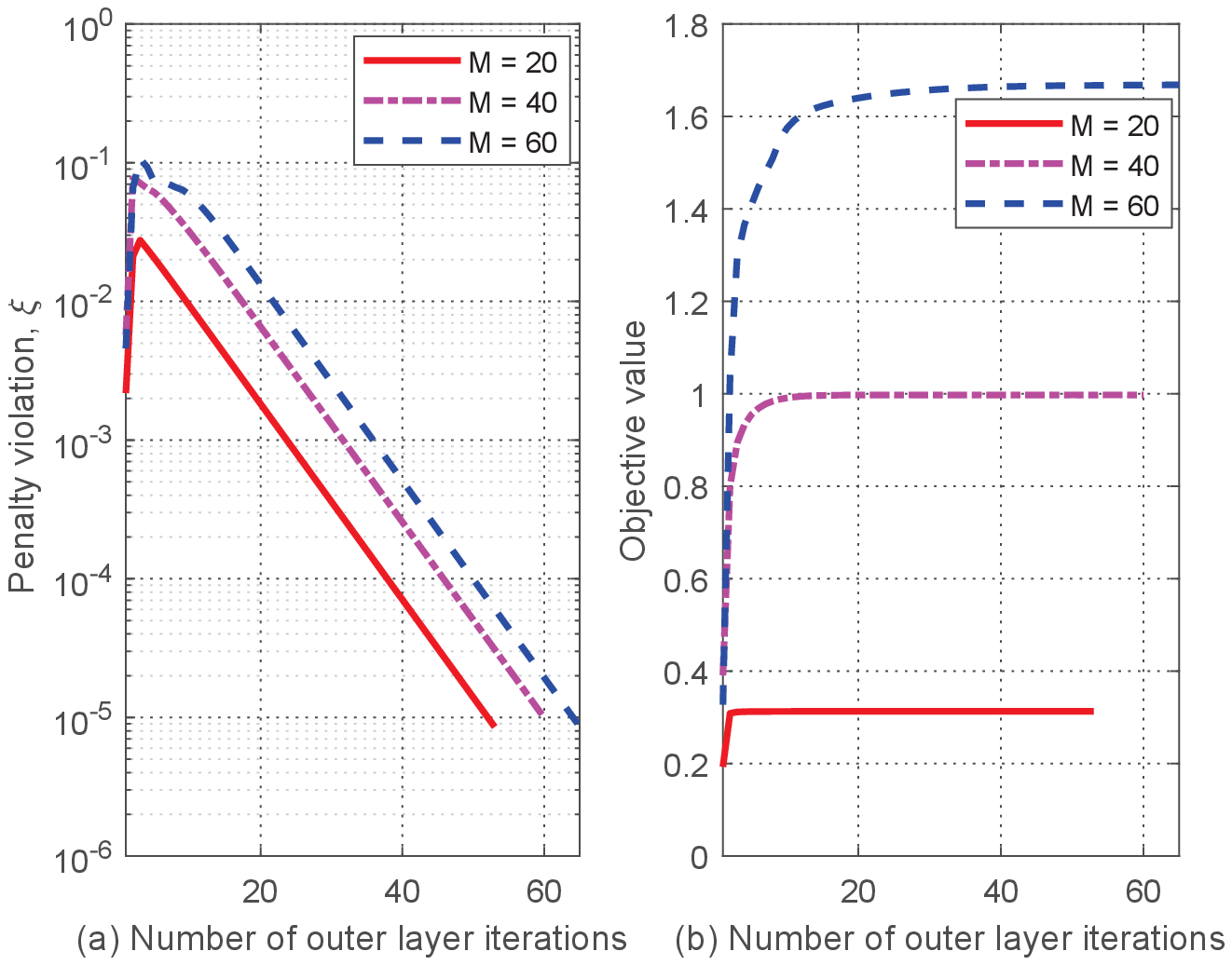}
		\caption{One snapshot for showing convergence behaviour of Algorithm~\ref{alg2}.}\label{fig_convergence}
	\end{minipage}
	\hspace{10pt}
	\begin{minipage}[t]{0.45\textwidth}
		\centering
		\includegraphics[width=2.8in]{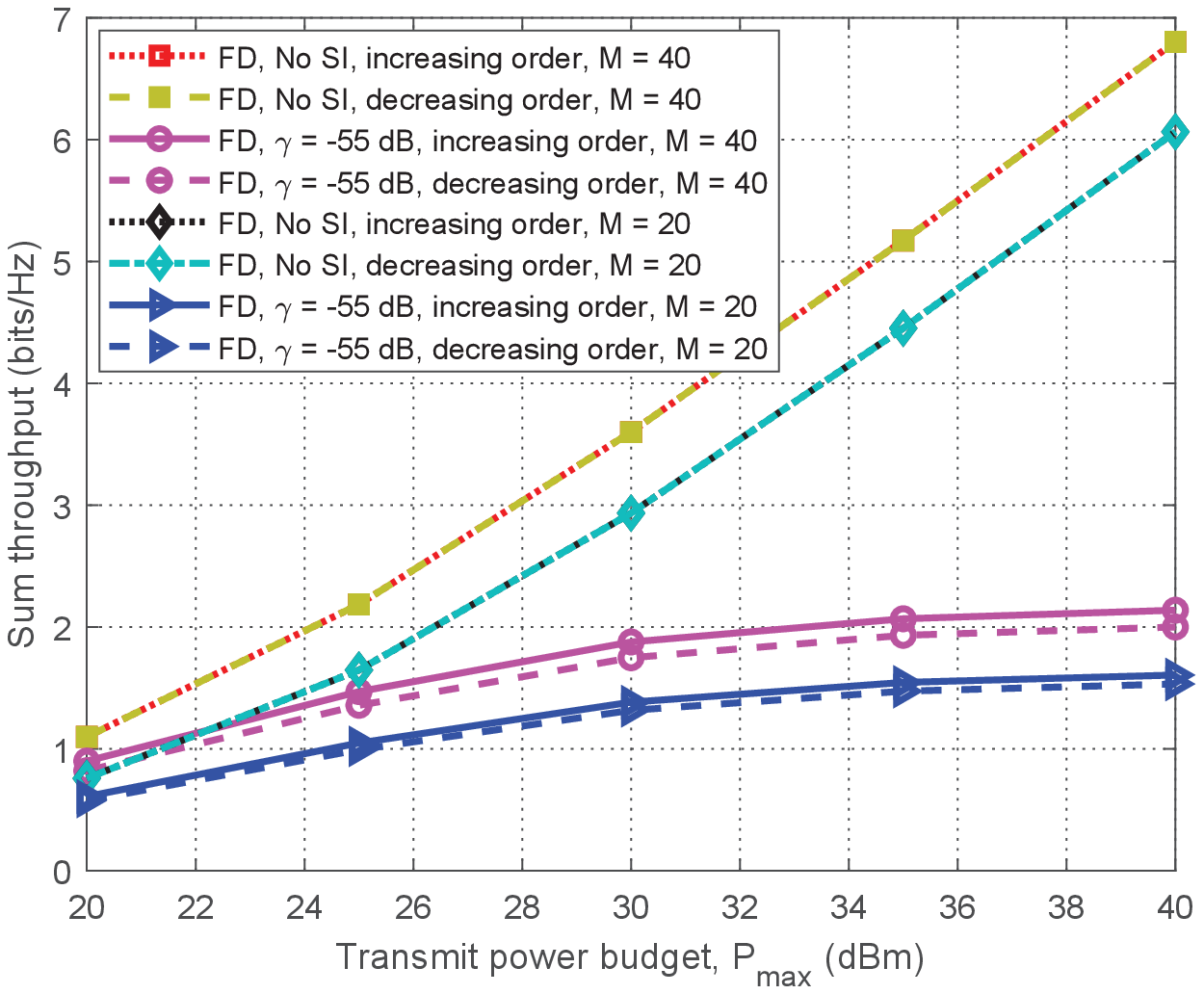}
		\caption{Sum throughput versus  $P_{\max}$  for different scheduling orders and  $M$ with $K=10$.}\label{fig3}
	\end{minipage}
\vspace{-15pt} 
\end{figure}
\subsection{Fully Dynamic IRS Beamforming}
In this subsection, we study fully dynamic IRS beamforming and compare the IRS-aided FD-WPCN with several benchmarks   in terms of throughput under different setups such as  device scheduling, transmit power $P$ and  the number of IRS reflecting elements $M$.

Before discussing the system performance  of the IRS-aided WPCN, we first verify  the effectiveness of penalty-based Algorithm~\ref{alg2} for the dynamic IRS beamforming with  imperfect SIC.  The constraint violation and convergence behaviour of Algorithm~\ref{alg2} for  one snapshot with $P_{\max}=30~{\rm dBm}$  with the  different number of IRS reflecting elements $M$, namely, $M=20$, $M=40$, and $M=60$,  are  plotted in Fig.~\ref{fig_convergence}. From Fig.~\ref{fig_convergence}(a), it  is observed that constraint violation $\xi$  converges very fast with the value decreasing to  predefined accuracy $10^{-5}$ after about  $53$ iterations for $M=20$, which indicates that the equality constraint  \eqref{p1_1_const1}  in problem \eqref{p1_1} can be eventually satisfied. Even for $M=60$, only about $65$ iterations are required for reaching the violation predefined accuracy, which again demonstrates its effectiveness. Note that there exists fluctuations of the penalty violation  in
the initial few iterations. This is mainly because when the initial
penalty $\rho$ is relatively large, the solution obtained by the penalty-based algorithm does not satisfy  the equality in \eqref{p1_1_const1}.  While as   $\rho$  decreases with the iteration number increases, the penalty  violation $\mathop {\max }\limits_{k \in {{\cal K}_2}} \left\{ {\left| {{z_k} - \Gamma \left( {\beta \gamma {P_k}/{\sigma ^2} + 1} \right)} \right|} \right\}$ is forced to approach the predefined
accuracy $\varepsilon_2=10^{-5}$. As such, the penalty-based algorithm is guaranteed to
converge finally. This can also be observed more clearly in Fig.~\ref{fig_convergence}(b),  where 
the penalized objective values of \eqref{p1_1_penalty} obtained by  different $M$ all  increase quickly with the number of iterations and finally converge.
\begin{figure}[!t]
	\centering
	%\vspace{-20pt}
	\begin{minipage}[t]{0.49\textwidth}
		\centering
		\includegraphics[width=3in]{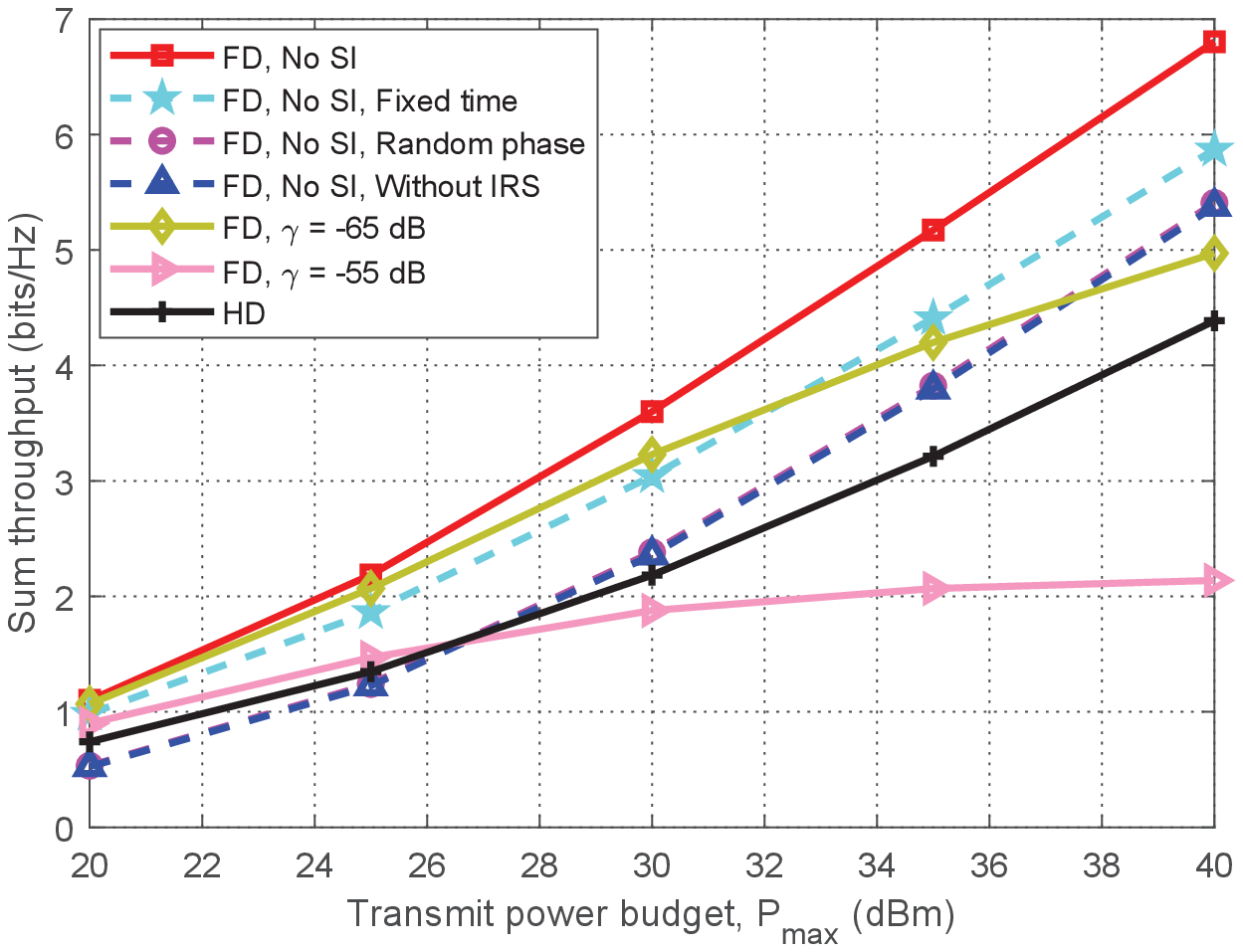}
		\caption{Sum throughput versus  $P_{\max}$ with $K=10$ and $M=40$.}\label{fig4} 
	\end{minipage}
\hspace{10pt}
	\begin{minipage}[t]{0.45\textwidth}
	\centering
	\includegraphics[width=2.8in]{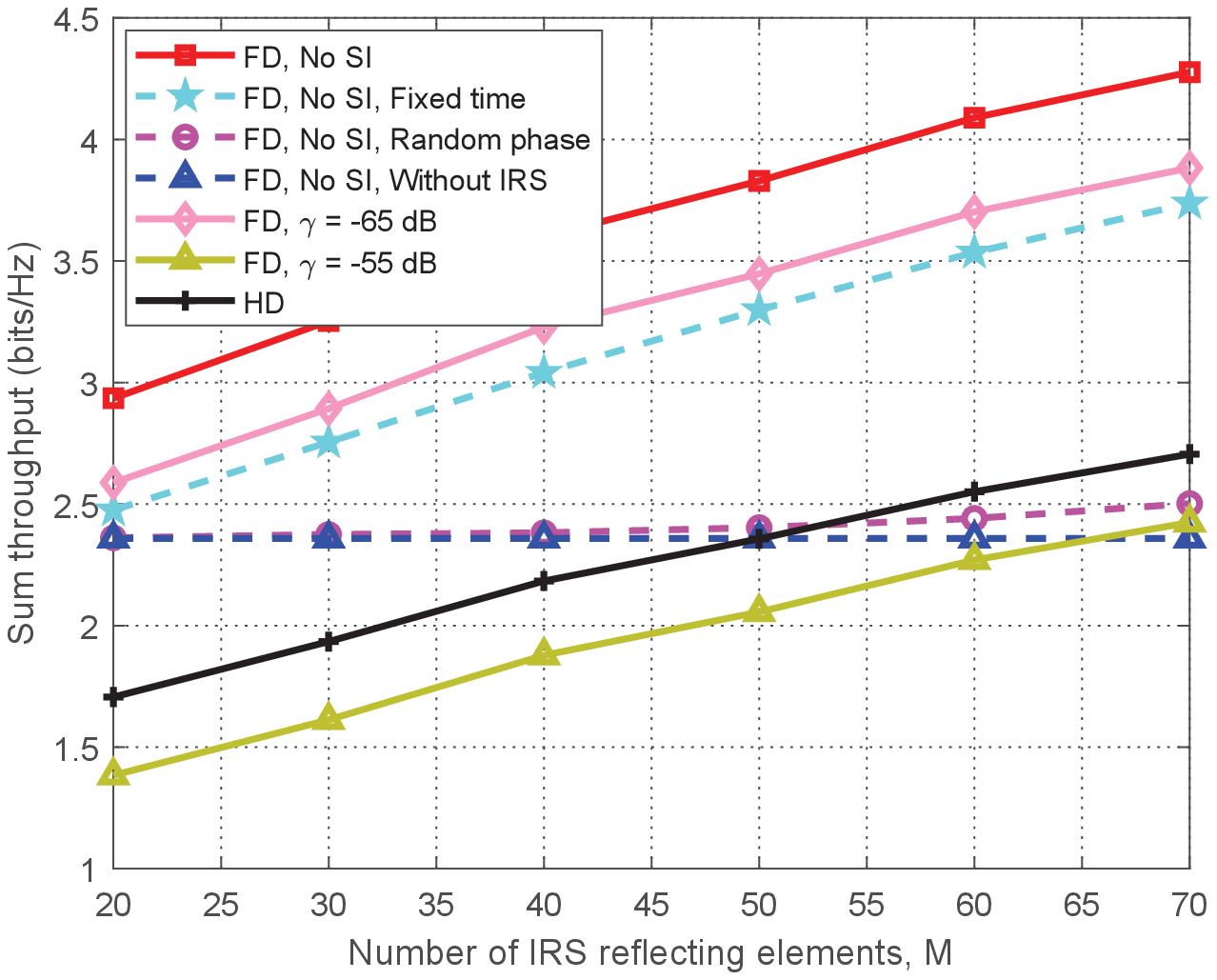}
	\caption{Sum throughput versus   $M$.}\label{fig5}
\end{minipage}
\vspace{-15pt} 
\end{figure}
 \subsubsection{Effect of   device scheduling} In Fig.~\ref{fig3}, we study the impact of device scheduling on the sum throughput of the  IRS-aided FD-WPCN.  We consider two scheduling schemes: (a) increasing order of SNR, where the device with the highest SNR of the HAP-device link is scheduled to transmit first; (b) decreasing order of SNR, where the device with the lowest SNR of the HAP-device link is scheduled to transmit first. In addition, we also study  the effects of scheduling under different setups, such as SI $\gamma$ and the number of reflecting elements $M$, on the system performance. It is  observed from Fig.~\ref{fig3}  that the sum throughput obtained by all 
schemes increases when the  HAP transmit power budget $P_{\max}$ increases.  The performance gap between increasing order and decreasing order for perfect SIC case is   negligible, even $M$ is large. For imperfect SIC,  the increasing order scheduling scheme  still achieves a negligible  sum throughput compared to  the decreasing order scheduling scheme, even for $M=40$. This indicates that the scheduling order is not significant  in the considered system model.   As such,   we adopt the increasing order scheduling as our scheduling strategy in the following  simulations.

 \subsubsection{Effect of   HAP transmit power}  In Fig.~\ref{fig4},  we compare the sum throughput obtained by all schemes versus  $P_{\max}$ with $K=10$ and $M=40$. In order to show the performance  of  IRS-aided FD-WPCN, we adopt the following schemes for comparison: (a) Perfect SIC: we study four cases, namely, ``FD, No SI'', 
 ``FD, No SI, Fixed time'', ``FD, No SI, Random phase'', and ``FD, No SI, without IRS''. For ``FD, No SI'', this is our proposed scheme by jointly optimizing the  phase shifts and the time allocation, which  is  solved by Algorithm~\ref{alg1}; For ``FD, No SI, Fixed time'', we only optimize the phase shifts with the time equally allocated to each time slot; For ``FD, No SI, Random phase'', we only optimize the time allocation with the phase shift  at each element randomly following  $\left[ {0,2\pi } \right)$ for each channel realization;  For ``FD, No SI, Without IRS'', we only optimize the  time allocation without using the IRS.
 (b) Imperfect SIC: we consider two cases, namely, $\gamma=-55 ~{\rm dB}$ and $\gamma=-65 ~{\rm dB}$. (c) HD, we consider the  sum  throughput optimization problem  for the IRS-aided HD-WPCN with the harvest-then-transmit protocol proposed in  \cite{Ju2014Throughput}, by jointly the  phase shifts and the time allocation.
 
  It is observed  from Fig.~\ref{fig4} that the sum throughput obtained by all schemes increases with $P_{\max}$. For perfect SIC case, our proposed scheme outperforms than other schemes, which demonstrates  the benefits of joint optimization of phase shifts and time allocation. In addition, it can be seen that the  IRS with random phase achieves  nearly the  same performance as the case  without IRS since  the reflected signal is dissipated by the random phase  adjustment at the   IRS.
  Besides, for perfect SIC and  imperfect SIC with   $\gamma=-65 ~{\rm dB}$, both of them   outperform the IRS-aided HD-WPCN, which shows the superiority of IRS-aided FD-WPCN over the IRS-aided HD-WPCN if the SI is well suppressed. When $\gamma=-55 ~{\rm dB}$,   the IRS-aided FD-WPCN  outperforms the IRS-aided HD-WPCN only when $P_{\max}$ is  smaller than about $27~{\rm dBm}$, but performs worse   when  $P_{\max}$ becomes larger. This is because that as $P_{\max}$ increases,  a strong SI is imposed  on the HAP receiver side and thus will degrade the system performance.

 \begin{figure}[!t]
 	\centering
 	%	\hspace{10pt}
 	\begin{minipage}[t]{0.49\textwidth}
 		\centering
 		\includegraphics[width=2.8in]{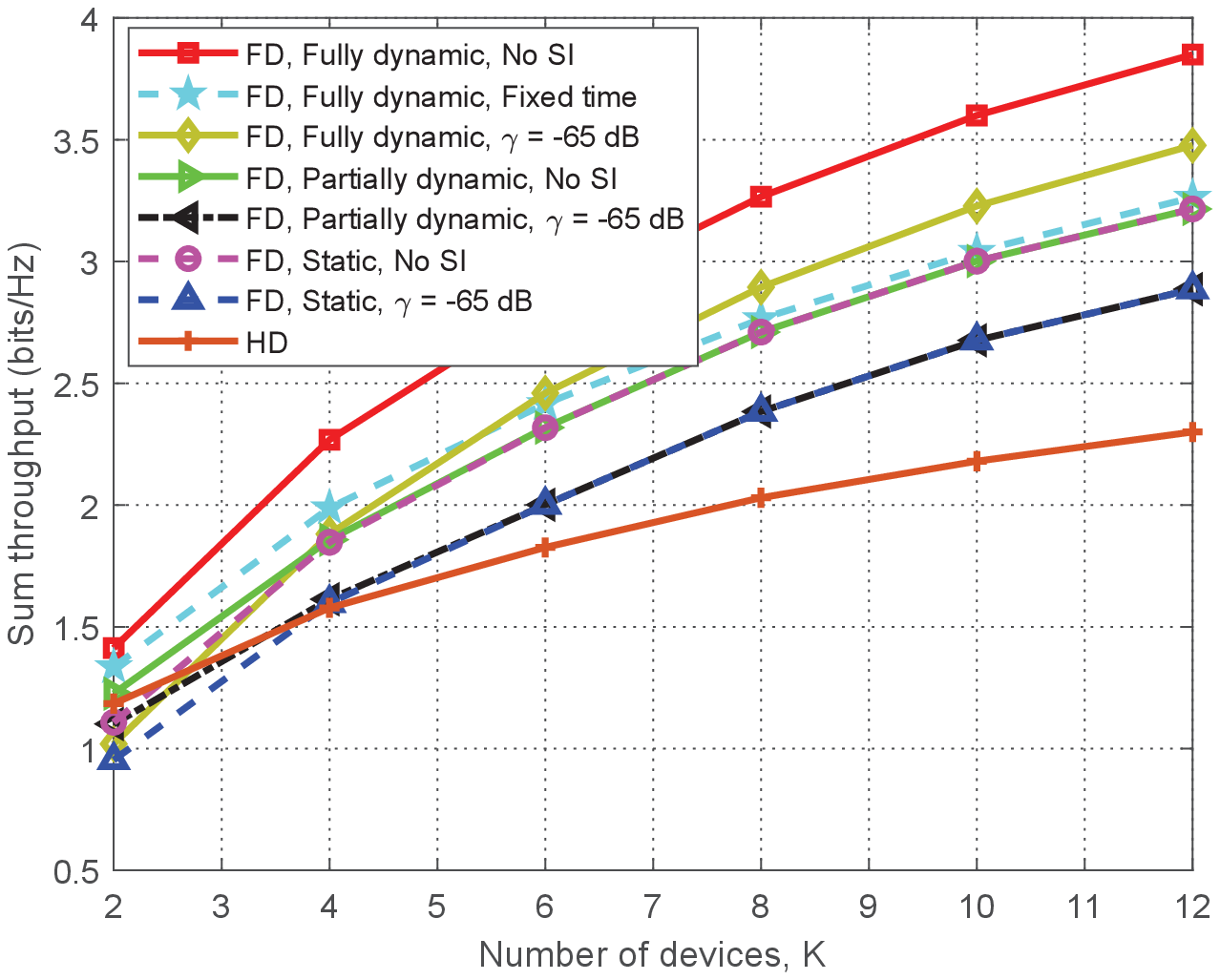}
 		\caption{Sum throughput versus   $K$ for three types of IRS configurations.}\label{fig6} 
 	\end{minipage}
 	\hspace{10pt}
  	\begin{minipage}[t]{0.45\textwidth}
 	\centering
 	\includegraphics[width=2.8in]{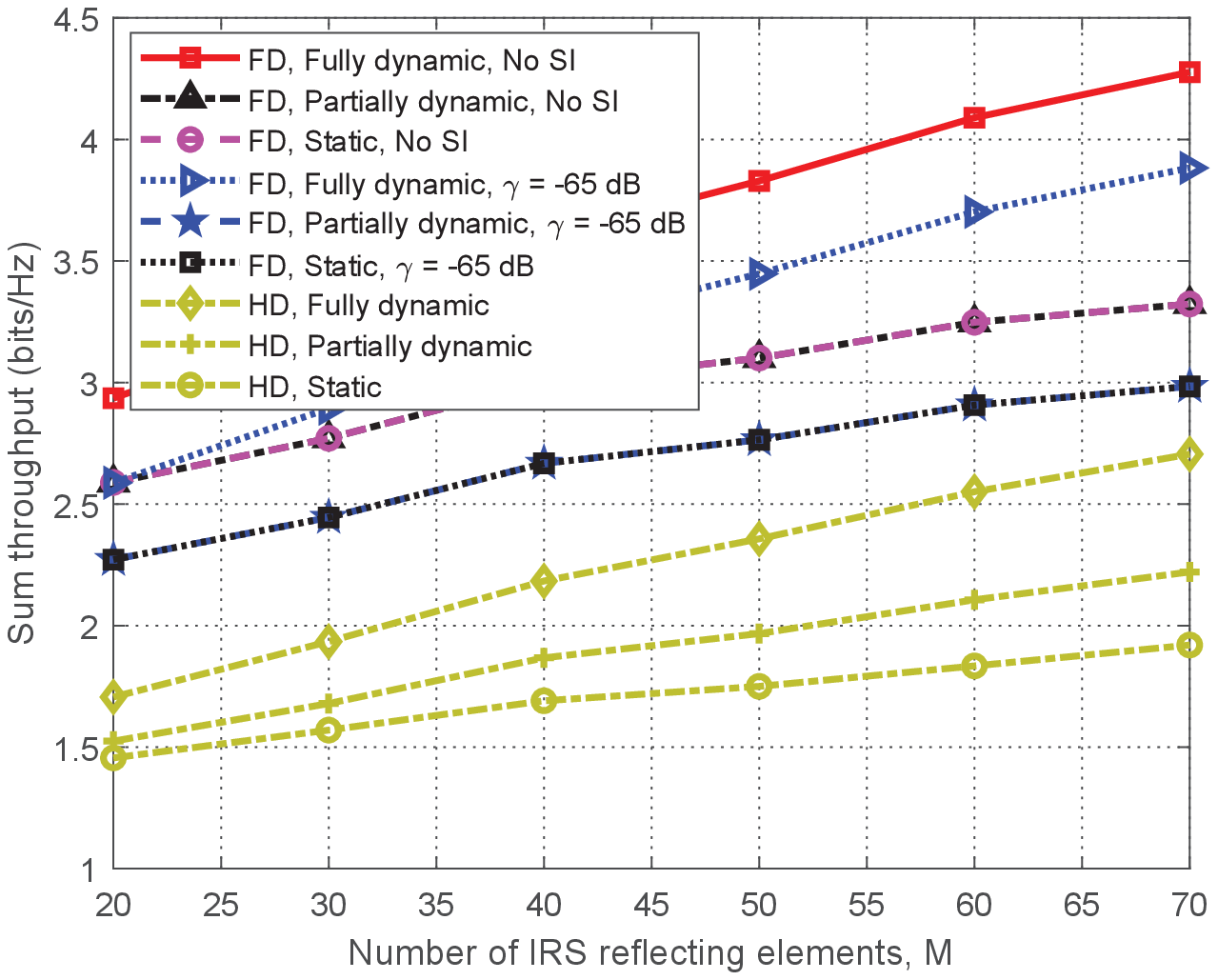}
 	\caption{Sum throughput versus $M$ for three types of IRS configurations  with $P_{\max}=30~{\rm dBm}$  and  $K=10$.}\label{fig8}
 \end{minipage}
\vspace{-15pt} 
\end{figure}
 
 \subsubsection{Effect of   number of reflecting elements} In  Fig.~\ref{fig5}, we compare the sum throughput obtained by all schemes versus   $M$ with  $P_{\max}=30~{\rm dBm}$  and  $K=10$.  It is observed that the sum throughput of the  schemes with IRS phase optimization monotonically increases with $M$ since more reflecting elements help achieve higher passive beamforming gain, which is  beneficial for both DL WET and UL WIT.  In addition, we can observe that for perfect SIC, the performance gap between the FD-WPCN with  IRS  and the FD-WPCN without IRS is magnified as $M$ increases. This is because the signal reflected by the IRS towards the desired devices becomes more focused   with increasing
 $M$.  In addition, the FD-WPCN with  random IRS phase shifts    outperforms that  without  IRS when  $M$ becomes large, since the IRS is
 able to reflect some of the dissipated signals back to the desired devices.
 Furthermore, it is observed that the SIC plays a key role in the system performance. For example, the  IRS-aided FD-WPCN with  $\gamma=-65 ~{\rm dB}$ significantly outperforms the IRS-aided HD-WPCN in terms of sum throughput, especially when $M$ becomes large, the performance gap will be more pronounced. However, when  $\gamma=-55 ~{\rm dB}$,  the achievable sum throughput  of  the IRS-aided FD-WPCN is smaller than  that of the IRS-aided HD-WPCN. This  indicates that the IRS-aided FD-WPCN  is superior over the  IRS-aided HD-WPCN  when the effective SIC can be implemented.

\subsection{Dynamic   versus Static IRS Beamforming }
In this subsection,  we compare three types of IRS beamforming configurations  in terms of  sum throughout. 
In  Fig.~\ref{fig6},  we study the effect of the number of devices $K$ with $P_{\max}=30~{\rm dBm}$  and  $M=40$ on the system performance for both IRS-aided  FD-WPCN and IRS-aided HD-WPCN. We can see that the sum throughput obtained by all the schemes increases as $K$ increases. This can be readily shown as follows. Denote 
	$\left\{ {\tau _k^{opt},k \in {{\cal K}_1}} \right\}$, $\left\{ {P_k^{opt},k \in {{\cal K}_1}} \right\}$, and  $\left\{ {{\bf{v}}_k^{opt},k \in {{\cal K}_1}} \right\}$ as  the optimal time allocation, transmit power, and phase shifts for $K$ devices. When the number of devices becomes $K+1$, $\left\{ {\tau _k^{opt} \cup \tau _{K + 1}^{opt},k \in {{\cal K}_1}} \right\}$, $\left\{ {P_k^{opt} \cup P_{K + 1}^{opt},k \in {{\cal K}_1}} \right\}$, and $\left\{ {{\bf{v}}_k^{opt} \cup {\bf{v}}_{K + 1}^{opt},k \in {{\cal K}_1}} \right\}$ with $\tau _{K + 1}^{opt} = P_{K + 1}^{opt} = 0$
	are the  feasible solutions   for the $K+1$ devices  and achieve  the same objective value compared to the  $K$ devices case. This shows that the objective value obtained with the $K$ devices case serves as a lower bound for that of the  case with $K+1$ devices, and thus a higher sum throughput is achieved when $K$ becomes large.
	
	In addition, several interesting insights are obtained. First, it is observed
	that for the perfect SIC, the   performance gap between our proposed scheme    and the fully dynamic IRS beamforming without time allocation optimization scheme  increases with the increasing of  $K$.  In particular, when $K=2$, the performance gap between two schemes is negligible,  while for $K=12$, the performance gap becomes significant. This indicates  that  the time allocation plays a important role when $K$ is large.
	Second,  if the SI is   well suppressed, e.g., $\gamma=-65 ~{\rm dB}$, the performance gap between HD and  FD  will be remarkably enlarged as $K$ becomes large. This implies that the effective SIC is indeed needed in   practical applications. Third, it is observed that for the perfect SIC case, when $K\le 4$,  the partially dynamic IRS beamforming scheme achieves only a slight higher performance gain than the static IRS beamforming scheme, while the performance gap is even negligible when  $K\geq 4$. This can be explained as follows. As $K$ is small, which indicates that the number of   time slots   for energy harvesting during UL WIT is small and the dedicated DL WET time slot plays a important role in improving system performance.  As such, an exclusive   IRS phase-shift vector  set for DL WET will significantly improve the harvested energy of the devices.  For the extreme case where $K=1$, the dedicated DL WET time slot must be required, since otherwise, no energy can be harvested. Similar  results can also be obtained for the case with   imperfect SIC.

In Fig.~\ref{fig8}, we study the sum throughput versus $M$ for the three types of the IRS  beamforming configurations with $P_{\max}=30~{\rm dBm}$  and  $K=10$. First, it is observed that for the  FD-WPCN with perfect SIC, the fully dynamic IRS beamforming case outperforms both partially dynamic and static IRS beamforming cases.  This is expected since for former, the IRS is able to proactively generate artificial time-varying channels over each time slot to  adapt to WET and WIT   so as to improve the system performance. We also find that the static IRS beamforming case is able to achieve  the same performance compared to  the partially dynamic  IRS beamforming case since the duration of DL WET time slot is almost zero if $P_{\max}$ is sufficiently large.  This conclusion is similar to the FD-WPCN without IRS derived  in  Corollary $3.1$ of \cite{ju2014optimal}. This unveils  us that  the dedicated DL WET stage  and  the DL  IRS phase shift optimization are not needed in practice if  $P_{\max}$ is sufficiently large.
For the   FD-WPCN with imperfect SIC, we can obtain   similar results. However, the results do not hold for the HD case. It is observed that the  HD with  partially dynamic  IRS beamforming   outperforms that with static IRS beamforming, and the performance gap becomes significant when $M$ is large. This can be explained as follows. Due to the  adopted harvest-then-transmit protocol for the  HD case,  the dedicated DL WET time slot must be needed, i..e, $\tau_0\ne0$, since otherwise, no energy can be used for UL WIT.  This  indicates that the IRS phase shift optimization  is beneficial for improving  DL WET. In addition, for the partially dynamic  IRS beamforming case, there are additional degrees of freedom of IRS phase shift optimization for UL WIT, thereby  improving the UL throughput. Finally, we can observe  that  the IRS-aided FD-WPCN is able to  achieve higher throughput compared to the  IRS-aided HD-WPCN, which again demonstrates the benefits of exploiting both  IRS and  FD in the WPCN.
\section{Conclusion}
In this paper, we proposed three types of IRS beamforming configurations for IRS-aided FD-WPCN, which strike a balance between the system performance and signaling overhead as well as implementation complexity. 
 The  sum throughput  maximization problems for the three cases are formulated by jointly optimizing  the time allocation, the HAP transmit power, and   IRS phase shifts. We first investigated the  fully dynamic IRS beamforming optimization with perfect SIC  and proposed an efficient AO algorithm.  Then, we   extended it to the  case with imperfect SIC and proposed a penalty-based algorithm. We also studied partially dynamic IRS beamforming and static IRS beamforming, and extended the proposed AO and penalty algorithms based on the  DC framework optimization and SCA techniques.
Simulation results  demonstrated the benefits of the    IRS    for enhancing the performance of the  FD-WPCN, especially when  the  fully dynamic IRS  beamforming is  adopted, and also showed that the  IRS-aided FD-WPCN is able to achieve significantly  performance gain compared to the  IRS-aided HD-WPCN when the SI is well suppressed. 
The results in this paper can be further extended by considering multiple IRSs, frequency-selective channel model, imperfect CSI, etc., which will be left as future work.

\bibliographystyle{IEEEtran}
\bibliography{Full_Duplex_IRS_WPCN}

\begin{thebibliography}{10}
\providecommand{\url}[1]{#1}
\csname url@samestyle\endcsname
\providecommand{\newblock}{\relax}
\providecommand{\bibinfo}[2]{#2}
\providecommand{\BIBentrySTDinterwordspacing}{\spaceskip=0pt\relax}
\providecommand{\BIBentryALTinterwordstretchfactor}{4}
\providecommand{\BIBentryALTinterwordspacing}{\spaceskip=\fontdimen2\font plus
\BIBentryALTinterwordstretchfactor\fontdimen3\font minus
  \fontdimen4\font\relax}
\providecommand{\BIBforeignlanguage}[2]{{%
\expandafter\ifx\csname l@#1\endcsname\relax
\typeout{** WARNING: IEEEtran.bst: No hyphenation pattern has been}%
\typeout{** loaded for the language `#1'. Using the pattern for}%
\typeout{** the default language instead.}%
\else
\language=\csname l@#1\endcsname
\fi
#2}}
\providecommand{\BIBdecl}{\relax}
\BIBdecl

\bibitem{lu2015Wireless}
X.~{Lu}, P.~{Wang}, D.~{Niyato}, D.~I. {Kim}, and Z.~{Han}, ``Wireless networks
  with {RF} energy harvesting: A contemporary survey,'' \emph{IEEE Commun.
  Surveys Tuts.}, vol.~17, no.~2, pp. 757--789, 2nd Quat., 2015.

\bibitem{wu2017an}
Q.~{Wu}, G.~Y. {Li}, W.~{Chen}, D.~W.~K. {Ng}, and R.~{Schober}, ``An overview
  of sustainable green {5G} networks,'' \emph{IEEE Wireless Commun.}, vol.~24,
  no.~4, pp. 72--80, Aug. 2017.

\bibitem{zhang2013mimo}
R.~{Zhang} and C.~K. {Ho}, ``{MIMO} broadcasting for simultaneous wireless
  information and power transfer,'' \emph{IEEE Trans. Wireless Commun.},
  vol.~12, no.~5, pp. 1989--2001, May 2013.

\bibitem{wang2018anewlook}
L.~{Wang}, K.~{Wong}, S.~{Jin}, G.~{Zheng}, and R.~W. {Heath}, ``A new look at
  physical layer security, caching, and wireless energy harvesting for
  heterogeneous ultra-dense networks,'' \emph{IEEE Commun. Mag.}, vol.~56,
  no.~6, pp. 49--55, Jun. 2018.

\bibitem{kim2015Survey}
D.~{Kim}, H.~{Lee}, and D.~{Hong}, ``A survey of in-band full-duplex
  transmission: From the perspective of {PHY} and {MAC} layers,'' \emph{IEEE
  Commun. Surveys Tuts.}, vol.~17, no.~4, pp. 2017--2046, 4th Quat., 2015.

\bibitem{riihonen2011mitigation}
T.~Riihonen, S.~Werner, and R.~Wichman, ``Mitigation of loopback
  self-interference in full-duplex {MIMO} relays,'' \emph{IEEE Trans. Signal
  Process.}, vol.~59, no.~12, pp. 5983--5993, Dec. 2011.

\bibitem{duarte2010full}
M.~Duarte and A.~Sabharwal, ``Full-duplex wireless communications using
  off-the-shelf radios: Feasibility and first results,'' in \emph{Proc.
  Asilomar Conf. Signals, Syst., Comput.}, Nov. Pacific Grove, CA, USA, 2010,
  pp. 1558--1562.

\bibitem{zhang2015full}
Z.~{Zhang}, X.~{Chai}, K.~{Long}, A.~V. {Vasilakos}, and L.~{Hanzo}, ``Full
  duplex techniques for {5G} networks: self-interference cancellation, protocol
  design, and relay selection,'' \emph{IEEE Commun. Mag.}, vol.~53, no.~5, pp.
  128--137, May 2015.

\bibitem{zhou2017integrated}
J.~Zhou, N.~Reiskarimian, J.~Diakonikolas, T.~Dinc, T.~Chen, G.~Zussman, and
  H.~Krishnaswamy, ``Integrated full duplex radios,'' \emph{IEEE Commun. Mag.},
  vol.~55, no.~4, pp. 142--151, Apr. 2017.

\bibitem{wei2019resource}
Z.~{Wei}, S.~{Sun}, X.~{Zhu}, D.~{In Kim}, and D.~W.~K. {Ng}, ``Resource
  allocation for wireless-powered full-duplex relaying systems with nonlinear
  energy harvesting efficiency,'' \emph{IEEE Trans. Veh. Technol.}, vol.~68,
  no.~12, pp. 12\,079--12\,093, Dec. 2019.

\bibitem{ju2014optimal}
H.~{Ju} and R.~{Zhang}, ``Optimal resource allocation in full-duplex
  wireless-powered communication network,'' \emph{IEEE Trans. Commun.},
  vol.~62, no.~10, pp. 3528--3540, Oct. 2014.

\bibitem{kang2015full}
X.~{Kang}, C.~K. {Ho}, and S.~{Sun}, ``Full-duplex wireless-powered
  communication network with energy causality,'' \emph{IEEE Trans. Wireless
  Commun.}, vol.~14, no.~10, pp. 5539--5551, Oct. 2015.

\bibitem{leng2016multi}
S.~{Leng}, D.~W.~K. {Ng}, N.~{Zlatanov}, and R.~{Schober}, ``Multi-objective
  resource allocation in full-duplex {SWIPT} systems,'' in \emph{Proc. IEEE
  Int.Conf. Commun. (ICC),}, May Kuala Lumpur, Malaysia, 2016, pp. 1--7.

\bibitem{xu2018Hybrid}
K.~{Xu} \emph{et~al.}, ``Hybrid time-switching and power splitting {SWIPT} for
  full-duplex massive {MIMO} systems: A beam-domain approach,'' \emph{IEEE
  Trans. Veh. Technol.}, vol.~67, no.~8, pp. 7257--7274, Aug. 2018.

\bibitem{wu2020intelligentarxiv}
Q.~Wu, S.~Zhang, B.~Zheng, C.~You, and R.~Zhang, ``Intelligent reflecting
  surface aided wireless communications: A tutorial,'' \emph{IEEE Trans.
  Commun., to appear}, 2020.

\bibitem{marco2020smart}
{M. Di Renzo} \emph{et~al.}, ``Smart radio environments empowered by
  reconfigurable {AI} meta-surfaces: An idea whose time has come,''
  \emph{EURASIP J. Wireless Commun. Netw.}, vol. 2019, no.~1, pp. 1--20, May
  2019.

\bibitem{WU2020towards}
Q.~{Wu} and R.~{Zhang}, ``Towards smart and reconfigurable environment:
  Intelligent reflecting surface aided wireless network,'' \emph{IEEE Commun.
  Mag.}, vol.~58, no.~1, pp. 106--112, Jan. 2020.

\bibitem{huang2018Achievable}
C.~{Huang}, A.~{Zappone}, M.~{Debbah}, and C.~{Yuen}, ``Achievable rate
  maximization by passive intelligent mirrors,'' in \emph{Proc. IEEE ICASSP},
  Apr. Calgary, AB, Canada, 2018, pp. 3714--3718.

\bibitem{wu2019intelligentxx}
Q.~Wu and R.~Zhang, ``Intelligent reflecting surface enhanced wireless network
  via joint active and passive beamforming,'' \emph{IEEE Trans. Wireless
  Commun.}, vol.~18, no.~11, pp. 5394--5409, Nov. 2019.

\bibitem{hu2021robust}
\BIBentryALTinterwordspacing
S.~Hu, Z.~Wei, Y.~Cai, C.~Liu, D.~W.~K. Ng, and J.~Yuan, ``Robust and secure
  sum-rate maximization for multiuser {MISO} downlink systems with
  self-sustainable {IRS},'' 2021. [Online]. Available:
  \url{https://arxiv.org/abs/2101.10549.}
\BIBentrySTDinterwordspacing

\bibitem{tang2021Wireless}
W.~{Tang} \emph{et~al.}, ``Wireless communications with reconfigurable
  intelligent surface: Path loss modeling and experimental measurement,''
  \emph{IEEE Trans. Wireless Commun.}, vol.~20, no.~1, pp. 421--439, Jan. 2021.

\bibitem{Intelligent2020guan}
X.~{Guan}, Q.~{Wu}, and R.~{Zhang}, ``Intelligent reflecting surface assisted
  secrecy communication: Is artificial noise helpful or not?'' \emph{IEEE
  Wireless Commun. Lett.}, vol.~9, no.~6, pp. 778--782, Jun. 2020.

\bibitem{zhang2020robust}
Z.~{Zhang}, L.~{Lv}, Q.~{Wu}, H.~{Deng}, and J.~{Chen}, ``Robust and secure
  communications in intelligent reflecting surface assisted {NOMA} networks,''
  \emph{IEEE Commun. Lett.}, 2020, early access, doi:
  10.1109/LCOMM.2020.3039811.

\bibitem{yu2020robust}
X.~{Yu}, D.~{Xu}, Y.~{Sun}, D.~W.~K. {Ng}, and R.~{Schober}, ``Robust and
  secure wireless communications via intelligent reflecting surfaces,''
  \emph{IEEE J. Sel. Areas Commun.}, vol.~38, no.~11, pp. 2637--2652, Nov.
  2020.

\bibitem{pan2020multicell}
C.~{Pan}, H.~{Ren}, K.~{Wang}, W.~{Xu}, M.~{Elkashlan}, A.~{Nallanathan}, and
  L.~{Hanzo}, ``Multicell {MIMO} communications relying on intelligent
  reflecting surfaces,'' \emph{IEEE Trans. Wireless Commun.}, vol.~19, no.~8,
  pp. 5218--5233, Aug. 2020.

\bibitem{hua2020intelligent}
M.~Hua, Q.~Wu, D.~W.~K. Ng, J.~Zhao, and L.~Yang, ``Intelligent reflecting
  surface-aided joint processing coordinated multipoint transmission,''
  \emph{IEEE Trans. Commun.}, 2020, early access, doi:
  10.1109/TCOMM.2020.3042275.

\bibitem{xie2020max}
H.~{Xie}, J.~{Xu}, and Y.~F. {Liu}, ``Max-min fairness in {IRS}-aided
  multi-cell {MISO} systems with joint transmit and reflective beamforming,''
  \emph{IEEE Trans. Wireless Commun.}, 2020, early access, doi:
  10.1109/TWC.2020.3033332.

\bibitem{hui2019Reflections}
\BIBentryALTinterwordspacing
H.~Long \emph{et~al.}, ``Reflections in the sky: Joint trajectory and passive
  beamforming design for secure {UAV} networks with reconfigurable intelligent
  surface,'' 2020. [Online]. Available: \url{https://arxiv.org/abs/2005.10559.}
\BIBentrySTDinterwordspacing

\bibitem{sixian2020robust}
\BIBentryALTinterwordspacing
S.~Li, B.~Duo, D.~R. Marco, M.~Tao, and X.~Yuan, ``Robust secure {UAV}
  communications with the aid of reconfigurable intelligent surfaces,'' 2020.
  [Online]. Available: \url{https://arxiv.org/abs/2008.09404.}
\BIBentrySTDinterwordspacing

\bibitem{wu2019weighted}
Q.~{Wu} and R.~{Zhang}, ``Weighted sum power maximization for intelligent
  reflecting surface aided {SWIPT},'' \emph{IEEE Wireless Commun. Lett.},
  vol.~9, no.~5, pp. 586--590, May 2020.

\bibitem{pan2019intelligent}
C.~Pan, H.~Ren, K.~Wang, M.~Elkashlan, A.~Nallanathan, J.~Wang, and L.~Hanzo,
  ``Intelligent reflecting surface enhanced {MIMO} broadcasting for
  simultaneous wireless information and power transfer,'' \emph{IEEE J. Sel.
  Areas Commun.}, vol.~38, no.~8, pp. 1719--1734, Aug. 2020.

\bibitem{li2020joint}
\BIBentryALTinterwordspacing
Z.~Li, W.~Chen, and Q.~Wu, ``Joint beamforming design and power splitting
  optimization in {IRS}-assisted {SWIPT} {NOMA} networks,'' 2020. [Online].
  Available: \url{https://arxiv.org/abs/2011.14778.}
\BIBentrySTDinterwordspacing

\bibitem{wu2020jointActive}
Q.~{Wu} and R.~{Zhang}, ``Joint active and passive beamforming optimization for
  intelligent reflecting surface assisted {SWIPT} under {QoS} constraints,''
  \emph{IEEE J. Sel. Areas Commun.}, vol.~38, no.~8, pp. 1735--1748, Aug. 2020.

\bibitem{wu2021intelligentoverview}
\BIBentryALTinterwordspacing
Q.~Wu, X.~Guan, and R.~Zhang, ``Intelligent reflecting surface aided wireless
  energy and information transmission: An overview,'' 2021. [Online].
  Available: \url{https://arxiv.org/abs/2106.07997v3.}
\BIBentrySTDinterwordspacing

\bibitem{zheng2020Intelligent}
Y.~{Zheng}, S.~{Bi}, Y.~J. {Zhang}, Z.~{Quan}, and H.~{Wang}, ``Intelligent
  reflecting surface enhanced user cooperation in wireless powered
  communication networks,'' \emph{IEEE Wireless Commun. Lett.}, vol.~9, no.~6,
  pp. 901--905, Jun. 2020.

\bibitem{zheng2020joint}
Y.~{Zheng}, S.~{Bi}, Y.~J.~A. {Zhang}, X.~{Lin}, and H.~{Wang}, ``Joint
  beamforming and power control for throughput maximization in {IRS}-assisted
  {MISO} {WPCNs},'' \emph{IEEE Internet of Things J.}, 2020, early access,doi:
  10.1109/JIOT.2020.3045703.

\bibitem{lyu2021OptimizedEnergy}
B.~{Lyu}, P.~{Ramezani}, D.~T. {Hoang}, S.~{Gong}, Z.~{Yang}, and
  A.~{Jamalipour}, ``Optimized energy and information relaying in
  self-sustainable {IRS}-empowered {WPCN},'' \emph{IEEE Trans. Commun.},
  vol.~69, no.~1, pp. 619--633, Jan. 2021.

\bibitem{rezaei2019Secrecy}
R.~{Rezaei}, S.~{Sun}, X.~{Kang}, Y.~L. {Guan}, and M.~R. {Pakravan}, ``Secrecy
  throughput maximization for full-duplex wireless powered {IoT} networks under
  fairness constraints,'' \emph{IEEE Internet of Things J.}, vol.~6, no.~4, pp.
  6964--6976, Aug. 2019.

\bibitem{iqbal2021minimum}
\BIBentryALTinterwordspacing
M.~S. Iqbal, Y.~Sadi, and S.~Coleri, ``Minimum length scheduling for multi-cell
  full duplex wireless powered communication networks,'' 2021. [Online].
  Available: \url{https://arxiv.org/abs/2101.08002.}
\BIBentrySTDinterwordspacing

\bibitem{wu2021irs}
\BIBentryALTinterwordspacing
Q.~Wu, X.~Zhou, W.~Chen, J.~Li, and X.~Zhang, ``{IRS}-aided {WPCNs}: A new
  optimization framework for dynamic {IRS} beamforming,'' 2021. [Online].
  Available: \url{https://arxiv.org/abs/2107.03251.}
\BIBentrySTDinterwordspacing

\bibitem{xu2020resource}
D.~{Xu}, X.~{Yu}, Y.~{Sun}, D.~W.~K. {Ng}, and R.~{Schober}, ``Resource
  allocation for {IRS}-assisted full-duplex cognitive radio systems,''
  \emph{IEEE Trans. Commun.}, vol.~68, no.~12, pp. 7376--7394, Dec. 2020.

\bibitem{shen2020beamforming}
H.~{Shen}, T.~{Ding}, W.~{Xu}, and C.~{Zhao}, ``Beamformig design with fast
  convergence for {IRS}-aided full-duplex communication,'' \emph{IEEE Commun.
  Lett.}, vol.~24, no.~12, pp. 2849--2853, Dec. 2020.

\bibitem{cai2020intelligent}
\BIBentryALTinterwordspacing
Y.~Cai, M.-M. Zhao, K.~Xu, and R.~Zhang, ``Intelligent reflecting surface aided
  full-duplex communication: Passive beamforming and deployment design,'' 2020.
  [Online]. Available: \url{https://arxiv.org/abs/2012.07218.}
\BIBentrySTDinterwordspacing

\bibitem{Ju2014Throughput}
H.~{Ju} and R.~{Zhang}, ``Throughput maximization in wireless powered
  communication networks,'' \emph{IEEE Trans. Wireless Commun.}, vol.~13,
  no.~1, pp. 418--428, Jan. 2014.

\bibitem{wqu2018Spectral}
Q.~{Wu}, W.~{Chen}, D.~W.~K. {Ng}, and R.~{Schober}, ``Spectral and
  energy-efficient wireless powered {IoT} networks: {NOMA} or {TDMA}?''
  \emph{IEEE Trans. Veh. Technol.}, vol.~67, no.~7, pp. 6663--6667, Jul. 2018.

\bibitem{Day2012Full}
B.~P. {Day}, A.~R. {Margetts}, D.~W. {Bliss}, and P.~{Schniter}, ``Full-duplex
  bidirectional {MIMO}: Achievable rates under limited dynamic range,''
  \emph{IEEE Trans. Signal Process.}, vol.~60, no.~7, pp. 3702--3713, Jul.
  2012.

\bibitem{boyd2004convex}
S.~Boyd and L.~Vandenberghe, \emph{Convex Optimization}.\hskip 1em plus 0.5em
  minus 0.4em\relax Cambridge university press, 2004.

\bibitem{gondzio1996computational}
J.~Gondzio and T.~Terlaky, ``A computational view of interior point methods,''
  \emph{Advances in linear and integer programming. Oxford Lecture Series in
  Mathematics and its Applications}, vol.~4, pp. 103--144, 1996.

\bibitem{shi2016joint}
Q.~{Shi}, M.~{Hong}, X.~{Gao}, E.~{Song}, Y.~{Cai}, and W.~{Xu}, ``Joint
  source-relay design for full-duplex {MIMO} {AF} relay systems,'' \emph{IEEE
  Trans. Signal Process.}, vol.~64, no.~23, pp. 6118--6131, Dec. 2016.

\bibitem{sidiropoulos2006transmit}
N.~D. Sidiropoulos, T.~N. Davidson, and Z.-Q. Luo, ``Transmit beamforming for
  physical-layer multicasting,'' \emph{IEEE Trans. Signal Process.}, vol.~54,
  no.~6, pp. 2239--2251, Jun. 2006.

\bibitem{tao1997convex}
P.~D. Tao and L.~T.~H. An, ``Convex analysis approach to {DC} programming:
  theory, algorithms and applications,'' \emph{Acta math. vietnamica}, vol.~22,
  no.~1, pp. 289--355, 1997.

\bibitem{Yang2020Federated}
K.~{Yang}, T.~{Jiang}, Y.~{Shi}, and Z.~{Ding}, ``Federated learning via
  over-the-air computation,'' \emph{IEEE Trans. Wireless Commun.}, vol.~19,
  no.~3, pp. 2022--2035, Mar. 2020.

\end{thebibliography}
\end{document}